\documentclass[aip,jcp,reprint]{revtex4-1} 
\usepackage{graphicx}

\usepackage{mathtools}
\usepackage{graphicx}   
\usepackage{color}
\usepackage{ulem}
\usepackage{caption}
\usepackage{subcaption}

\usepackage{array}
\usepackage{booktabs}
\usepackage{multirow}

\usepackage{booktabs}

\newcommand{\referee}{\textcolor{black} }

\begin{document}
\title{Predicting reaction coordinates in energy landscapes with diffusion anisotropy}
\author{Pratyush Tiwary}
 \affiliation{Department of Chemistry,
  Columbia University, New York 10027, USA.}
	
	\author{B. J. Berne}
	 \email{bb8@columbia.edu}   
	 \affiliation{Department of Chemistry, Columbia University, New York 10027, USA.}

	\date{\today}
	
	\begin{abstract}
We consider a range of model potentials with metastable states undergoing molecular dynamics coupled to a thermal bath in the high friction regime, and consider how the optimal reaction coordinate depends on the diffusion anisotropy. For this we use our recently proposed method ``Spectral gap optimization of order parameters (SGOOP)'' (Tiwary and Berne, Proc. Natl. Acad. Sci. 113 2839 2016). We show how available information about dynamical observables in addition to static information can be incorporated into SGOOP, which can then be used to accurately determine the ``best'' reaction coordinate for arbitrary anisotropies. We compare our results with transmission coefficient calculations and published benchmarks where applicable or available respectively.
\end{abstract}

	\maketitle
\section{Introduction}
\label{intro}
The notion that the potential energy surface (PES), or often the potential of mean force (PMF), governs the choice and the form of the reaction pathway is a central tenet in chemical physics. Given the PES (PMF), numerous very effective and successful techniques exist\cite{neb,fromatob,weinan2007} that can identify the minimum energy (free energy) pathway between a collection of metastable states, which is then taken to be the reaction pathway. The reaction coordinate (RC) is proverbially defined as an abstract low-dimensional coordinate that best captures progress along this reaction pathway. We define the RC more rigorously in Sec. \ref{rc_def}.

 However, such a reaction pathway and reaction coordinate derived from a purely static framework can be  significantly perturbed by the stochastic nature of the environment, and specifically by how the various degrees of freedom couple to the bath and relax in response to perturbations in it \cite{van1982reactive,szabo_anisotropic,hynes1985chemical,kl1989diffusion,
johnson2012characterization,peters2013reaction,echeverria2014concerted}. This coupling can be quantified through either a friction coefficient, or through its inverse, the diffusion constant. In the limit of strong coupling to the environment, one then expects that the diffusion will differ for different degrees of freedom, this anisotropy will manifest itself in the reaction pathway. A certain reactive mode derived from the PES/PMF might no longer be the path of least resistance between two metastable states, and one needs to consider the work done not just against the static forces from PES/PMF, but the dynamic effect of the environment. In fact, several workers have pointed out that the effect of diffusion anisotropy can be so strong as to lead to the phenomenon of ``saddle point avoidance''\cite{northrup1983saddle,berezhkovskii1989rate}, where the saddle points between metastable states could altogether be avoided while moving between them.  

Especially when one deals with free energies (as contrasted to potential energies) which are often defined as functions of qualitatively different collective variables, the diffusivities for the various coordinates will not be the same, and could possibly even be position and temperature dependent. For instance, for the very important problem of protein-ligand (un)binding\cite{copeland2015drug,trypsin,p38}, these variables could be ligand-host relative displacement, their individual conformations, their hydration states and other variables \cite{sgoop_fullerene,agmon1992diffusive}.  These variables can naturally be expected to possess very different diffusivities. For example, at low temperatures the protein fluctuations can be completely frozen out, with the  protein effectively trapped in one metastable conformation, while ligand unbinding still happens\cite{steinbach1991ligand}. In such cases, the free energy barrier is thus not the only determinant of the dynamics, and the RC which encapsulates the dominant slow fluctuations in the system will thus be a function of the diffusion anisotropy. In fact, while the free energy can be calculated as a function of any arbitrary order parameter, it is the most informative when expressed as a function of the optimal RC. 
This work is restricted to simple model potentials, as we are mainly interested in understanding how the optimal RC depends on diffusional anisotropy---a question pertinent to a range of very practical problems involving for example protein conformational transitions, drug un-binding and the efficiency of enhanced sampling.\cite{sgoop,sgoop_fullerene,fullerene} 

In the current work, we tackle the problem of how diffusion anisotropy affects the RC for different model potential energy landscapes with high and arbitrary number of energy barriers. To identify the RC and its dependence on diffusion anisotropy, we propose a practical numerical solution using our recent method ``Spectral gap optimization of order parameters (SGOOP)''.\cite{sgoop}  For several potential surfaces  we demonstrate how dynamical information can be incorporated into SGOOP which can then be used to predict the RC for arbitrary diffusion anisotropies. We look at three 2-d model potentials undergoing Langevin dynamics with different values of the diffusion anisotropy, and predict how the RC changes with anisotropy for these different cases. Where possible, we validate the SGOOP-predicted RC with published results which use different methods for RC calculation\cite{szabo_anisotropic}, and also with extensive calculations of the reactive flux and the transmission coefficients along different putative RCs.  The method SGOOP, which is described in Sec. \ref{theory}, considers the inverse problem of what can we infer about the RC given limited information about the static and dynamic behavior of the system. The dynamic information in SGOOP can be explicitly implemented if one knows the diffusion tensor, or through a Maximum Caliber framework\cite{caliber1,dixit2015inferring}, where one effectively fits the response function to the environment depending on known dynamical observables. These dynamical observables are clarified in Sec. \ref{szabo_maxcal}. We use both the methods and find excellent agreement between the results for all three potentials. Given the robustness and reliability of SGOOP shown in this work, we expect it will be very useful for complex systems with qualitatively different collective variables that respond differently to the environment.

\section{Theory}
\label{theory}

\subsection{Reaction coordinate}
\label{rc_def}
In spite, and possibly due to, its ubiquity across the chemical kinetics literature, the reaction coordinate (RC) has multiple definitions, often depending on the intended use. Here we clarify our interpretation of the RC. For a given multidimensional complex system undergoing a certain dynamics, we define RC as a low-dimensional (say one or two dimensional) variable such that the multidimensional dynamics of the full system in terms of movement between different metastable states, can be ``best'' mapped into Markovian dynamics between various metastable states viewed as a function of the low-dimensional RC. In other words, the RC should have two features: (1) it should be able to demarcate between the various important metastable states that the full system possesses, and (2) there should be a separation of timescales between the relaxation times in the various metastable states, and the relatively slow transition times between them.  In Section \ref{rc_dynamics} we define the timescale separation more carefully through the notion of the spectral gap\cite{sgoop,diffusionmap,coifman,coifman2008diffusion}.

\subsection{Potential of mean force along putative reaction coordinate} 
\label{rc_pmf}
Without loss of generality, we consider a 2-dimensional energy landscape, in the space of Cartesian or generalized variables $x$ and $y$, with corresponding PES or PMF respectively given by $U(x,y)$. On this landscape we consider a putative RC $\hat{u} = \hat{x} cos(\theta)+\hat{y} sin(\theta)$, oriented at an angle $\theta$ from $\hat{x}$. For example, in Fig. \ref{fig:BS}(a) the magenta colored dashed line denotes a putative RC rotated by an angle $\theta$ from the x-axis (solid black line). Note that the relevant parameter for $\hat{u}$ parametrized as such is its direction, and it is invariant to translation. This could easily be generalized to more than 2 variables with the use of more than 1 angle. The PMF $F(q)$ where $q$ measures distance along the unit vector $\hat{u}$ i.e. $\vec{q} = q\hat{u}$ can be defined as
\begin{equation}
\label{PMF}
e^{-\beta F(q)} = {\int\int dx dy\delta(q-q(x,y))e^{-\beta U(x,y)}  \over \int\int dx dy e^{-\beta U(x,y)}   }
\end{equation}
where $\beta = {( k_B T)^{-1}}$ is the inverse temperature. The normalization in the denominator allows us to define an associated stationary probability density $p^0(q) = e^{-\beta F(q)} $. 

The PMF can be calculated analytically/numerically for different choices of $\hat{u}$ for the model potentials in this work. For more complex higher dimensional potentials it can be calculated through a range of sampling techniques as was shown in Ref.\cite{sgoop,sgoop_fullerene}. In this work, the primary interest is to see how the dynamical effects of diffusion anisotropy can be incorporated into SGOOP, and as such as far as the PMF is concerned, we evaluate it analytically/numerically. We found no perceivable differences in the results by using a numerically estimated PMF. However, as explained later, we calculate dynamical observables analytically as well as through sampling.

\subsection{Dynamics and spectral gap along putative reaction coordinate} 
\label{rc_dynamics}
Along a putative RC $\hat{u} = \hat{x} cos(\theta)+\hat{y} sin(\theta)$, we assume that the time-dependent probability density $p(q,t)$ is governed by a Smoluchowski Equation with position dependent diffusivity:
\begin{equation}
\label{smol}
{\partial p(q,t) \over \partial t} = {\partial \over \partial q}D(q)e^{-\beta F(q)} {\partial \over \partial q}e^{\beta F(q)} p(q,t) \equiv -Hp
\end{equation}

Under fairly general conditions, the operator $H$ has a discrete spectrum of non-negative eigenvalues $\{\lambda_i\}$ with $\lambda_0=0<\lambda_1 \leq \lambda_2 \leq ...$, and corresponding eigenvectors $\phi_i(q)$, and the general solution of Eq. \ref{smol} is given by
\begin{equation}
\label{smol_soln}
p(q,t) = p^0(q) + \Sigma_{i=1}^{\infty}c_i \phi_i(q) e^{-\lambda_i t}
\end{equation}

For a system with $n \geq1 $ slow processes (for example corresponding to existence of $n$ metastable states), the eigenspectrum of the operator $H$ will have a gap between the eigenvalues $\lambda_n$ and $\lambda_{n+1}$, i.e.  $\lambda_{n} \ll \lambda_{n+1}$. Please note that strictly speaking we consider the spectrum of eigenvalues of the associated master equation described in the following section which has similar properties. We refer to the quantity $e^{-\lambda_{n}}-e^{-\lambda_{n+1}}$ as the spectral gap. The larger the spectral gap, the larger will be the timescale separation between the slow and fast processes when the dynamics is viewed as a function of the putative RC $\hat{u}$. 

We consider a general diffusivity tensor in the Cartesian basis given by 
\[ \left( \begin{array}{cc}
D_x & 0\\
0 & D_y \end{array} \right)\] 
i.e. one without any off-diagonal terms.  Interesting complications would naturally arise if these cross-terms were not 0, but they can always be dealt with by diagonalizing the diffusion tensor and working with a different PES/PMF. We define the anisotropy parameter as $\delta = {D_y / D_x}$. We consider different values of $\delta$ ranging from $\delta \ll 1$ to $\delta \gg 1$, where for instance $\delta \ll 1$ represents the case when diffusion along $y$-axis is much slower than diffusion along the $x$-axis.

For simplicity, we also assume that the terms $D_{x}$ and $D_{y}$ have no positional dependence. Under this assumption, the diffusivity $D_{\theta}$ along any putative RC $\hat{u}$ rotated at angle $\theta$ from the $x-$axis can be removed from inside the partial derivative in Eq. \ref{smol}, which can now be written as:
\begin{equation}
\label{smol1}
{\partial p(q,t) \over \partial t} = D_{\theta} {\partial \over \partial q}e^{-\beta F(q)} {\partial \over \partial q}e^{\beta F(q)} p(q,t) 
\end{equation}

In fact for the two-dimensional case given the diffusivity tensor above we can easily calculate $D_{\theta}$:
\begin{equation}
\label{dtheta}
D_{\theta} = D_x cos^2(\theta) +  D_y sin^2(\theta)
\end{equation}

\subsection{Master equation along putative reaction coordinate}
\label{master}
We now spatially discretize the putative RC $\hat{u}$ by defining a grid $\{q_n\}$ along it, where $n$ takes integral values. Let $p_n(t)$ denote the instantaneous probability of being on any of these grid points $n$ at time $t$ (also discretized in intervals $\Delta t$), and $k_{mn}$ be the time-independent rate of transition from grid point $m$ to $n$ per unit time $\Delta t$. We can then write a master equation governing the flow of probability from one grid point to another:
\begin{equation}
{\Delta p_n (t)\over \Delta t } = \Sigma_m k_{mn} p_m (t) -  \Sigma_m 		
k_{nm} p_n (t)  \equiv \Sigma_m {\bf K}_{nm} p_m (t) 
\label{master}
\end{equation}

The matrix $\bf K$, where ${\bf K}_{nm}   = k_{mn}$, is the entirety of
all these rates. The diagonal terms in this matrix are obtained through $\Sigma_n k_{mn}=0$.
Under very general conditions the master equation possesses a unique stationary state solution, which we denote $p^0_n$ in accordance with Sec. \ref{rc_pmf}. We further assume that detailed balance is satisfied i.e. $p^0_nk_{nm} = p^0_mk_{mn} $ which can be shown to be true for closed isolated physical systems, but is in general a decent assumption.

Under these conditions, the master equation Eq.\ref{master} will also have a solution through expansion in eigenfunctions similar to Eq.\ref{smol_soln}. For convenience we use the same notation for the eigenfunction solution to the Smoluchowski equation as well as the master equation, and denote the eigenspectrum as $\{\lambda_i\}$ with $\lambda_0=0<\lambda_1 \leq \lambda_2 \leq ...$.
It is the spectral gap arising from this eigenspectrum that we use in SGOOP and in this work.

\subsection{Relations between transition rates and stationary probability}
\label{szabo_maxcal}
The transition rates $k_{mn}$ are underconstrained by the information presented so far including the detailed balance condition. We now introduce further relations connecting the transition rates $k_{mn}$ to the stationary probabilities $p^0_n$. Such a relation is fundamental to SGOOP, which we will summarize in the next section Sec. \ref{sgoop}. It can be derived in 2 ways which we describe here. In this work we use both these relations and assess how SGOOP performs with each of them.

The first way to connect the transition rates $k_{mn}$ to the stationary probabilities $p^0_n$ is by discretizing the Smoluchowski Equation (Eq. \ref{smol}) and then comparing it with the master equation (Eq. \ref{master}). This is described in Ref. \onlinecite{szabo_bicout} and below we give the final result:
\begin{equation}
k_{mn} = { D_\theta   \over 2d^2   } {\sqrt{   p^0_n \over p^0_m      }}
\label{rate_szabo}
\end{equation}
where $d = |u_k-u_{k-1}|$ $\forall k$ is the grid spacing, $D_\theta$ is the diffusivity along the putative RC $\hat{u}$, which as explained in Sec. \ref{rc_dynamics} we assume to depend on $\theta$ but not vary along $\hat{u}$. This approach is practical if one knows beforehand or calculates the diffusivity along various degrees of freedom for instance by using the approach of Ref. \cite{straub1990spatial,hummer2005position}, and then uses Eq. \ref{dtheta}.

A second approach connecting the transition rates $k_{mn}$ to the stationary probabilities $p^0_n$ is through the Maximum Caliber framework. While fairly general treatments are possible in this framework, here we start by assuming that apart from the stationary state information all one knows is the average value $\langle N\rangle$ of some dynamical observable $N$ over the full phase space. For instance $N$  could be the average number of first-nearest neighbor transitions observed in time $\Delta t$. Then $\langle N\rangle$ is given by
\begin{equation}
\langle N\rangle =\sum_{\substack{(m,n) \\ \forall |m-n|=1}} p^0_m k_{mn} N_{mn}=\sum_{\substack{(m,n) \\ \forall |m-n|=1}} p^0_m k_{mn}
\label{avgN}
\end{equation}
where we have used $ N_{mn} = 1$ $ \forall |m-n|=1$ and $0$ otherwise. We then seek an equation akin to Eq. \ref{rate_szabo} with the only known constraints being (1) principle of detailed balance (2) measurement of  $\langle N\rangle$. The principle of Maximum Caliber maximizes the path entropy subject to these constraints\cite{caliber1,dixit2015inferring}, and yields the same square-root of probability dependence, but now instead of using the explicit diffusivity, one works in the framework of Lagrange multipliers:
\begin{equation}
k_{mn} = \lambda   {\sqrt{   p^0_n \over p^0_m      }} = \left\{{\langle N \rangle \over{\sum \sqrt{p^0_n  p^0_m }  }}  \right\} {\sqrt{   p^0_n \over p^0_m      }}
\label{rate_maxcal}
\end{equation}
where we have calculated $\lambda$ by using the first equality of Eq. \ref{rate_maxcal} in Eq. \ref{avgN}. Note that similar expressions could be derived using a very wide range of dynamical observables and not just the one we use here.

With either Eq. \ref{rate_szabo} or \ref{rate_maxcal}, accompanied with either explicit diffusivities or measurement of $\langle N \rangle$ using molecular dynamics (MD) respectively, we can complement our knowledge of the stationary probability $p_0$  to calculate the full transition matrix $\bf K$, where ${\bf K}_{nm}   = k_{mn}$ as defined above. The eigenvalues of this matrix directly give us the spectral gap. 

A technical but crucial point must be mentioned here. In order to define and calculate $\langle N \rangle$ we have limited ourselves to first neighbors. We could go further than that, but then we subject ourselves to noisy and poorly sampled observations (i.e. transitions between far off grid points on $\{ u_k \}$). As such it is desirable to stick to first or first few neighbors in this context. However when we evaluate the eigenspectrum of the matrix $\bf K$,which could have come either  through Eq. \ref{rate_szabo} or Eq. \ref{rate_maxcal}, we are faced with a similar problem - how sparse should the matrix $\bf K$ be? Here, in contrast to the calculation of $\langle N \rangle$, it is desirable to have not just first neighbor terms as non-0 but ideally terms in the whole matrix. From practical perspectives for both these operations we start with first neighbor, move on to second neighbors, and so on, and stop when the spectral gap does not change. 

\begin{figure*}
  \centering
        \includegraphics[height=2.3in]{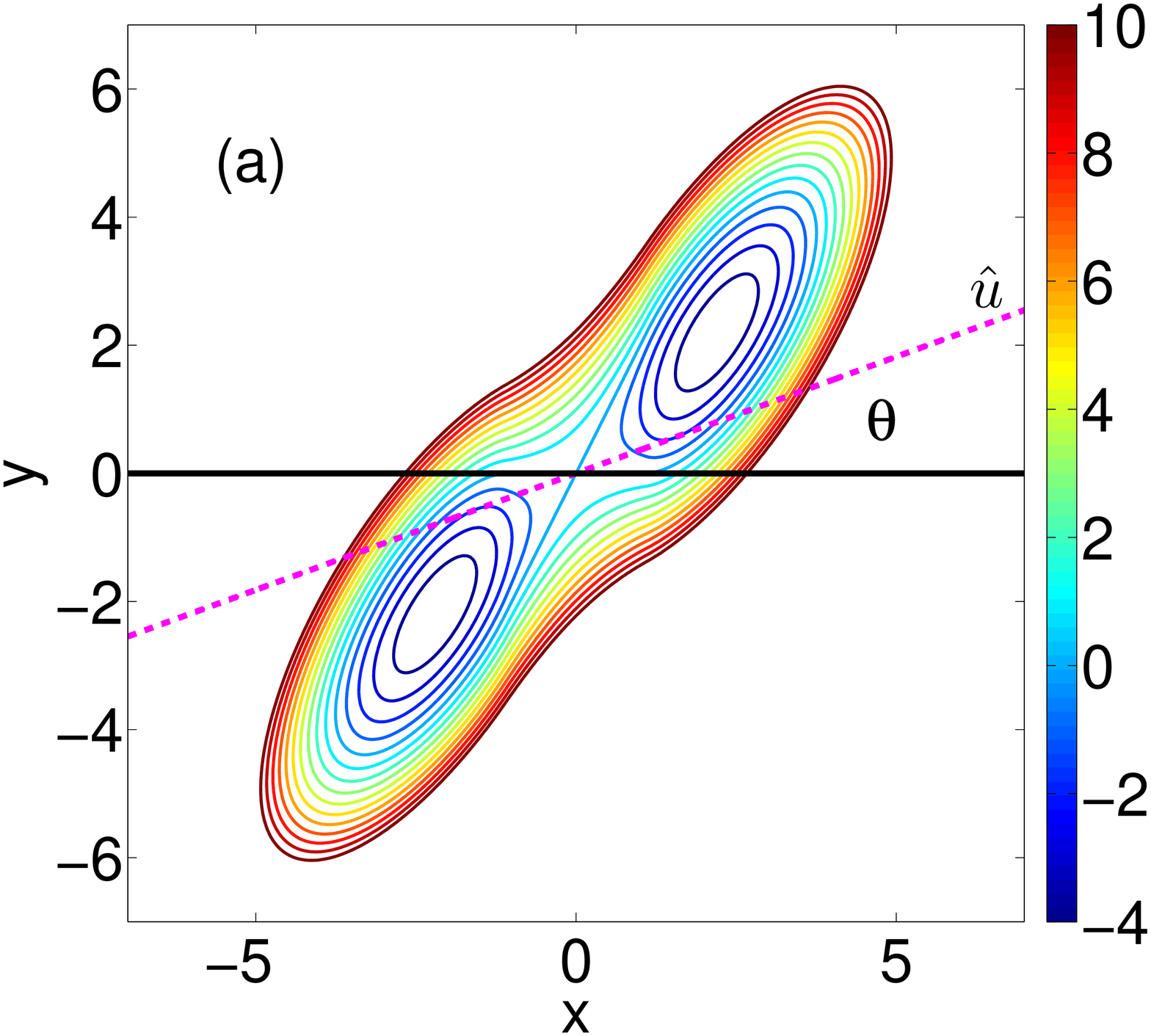} 
        \includegraphics[height=2.3in]{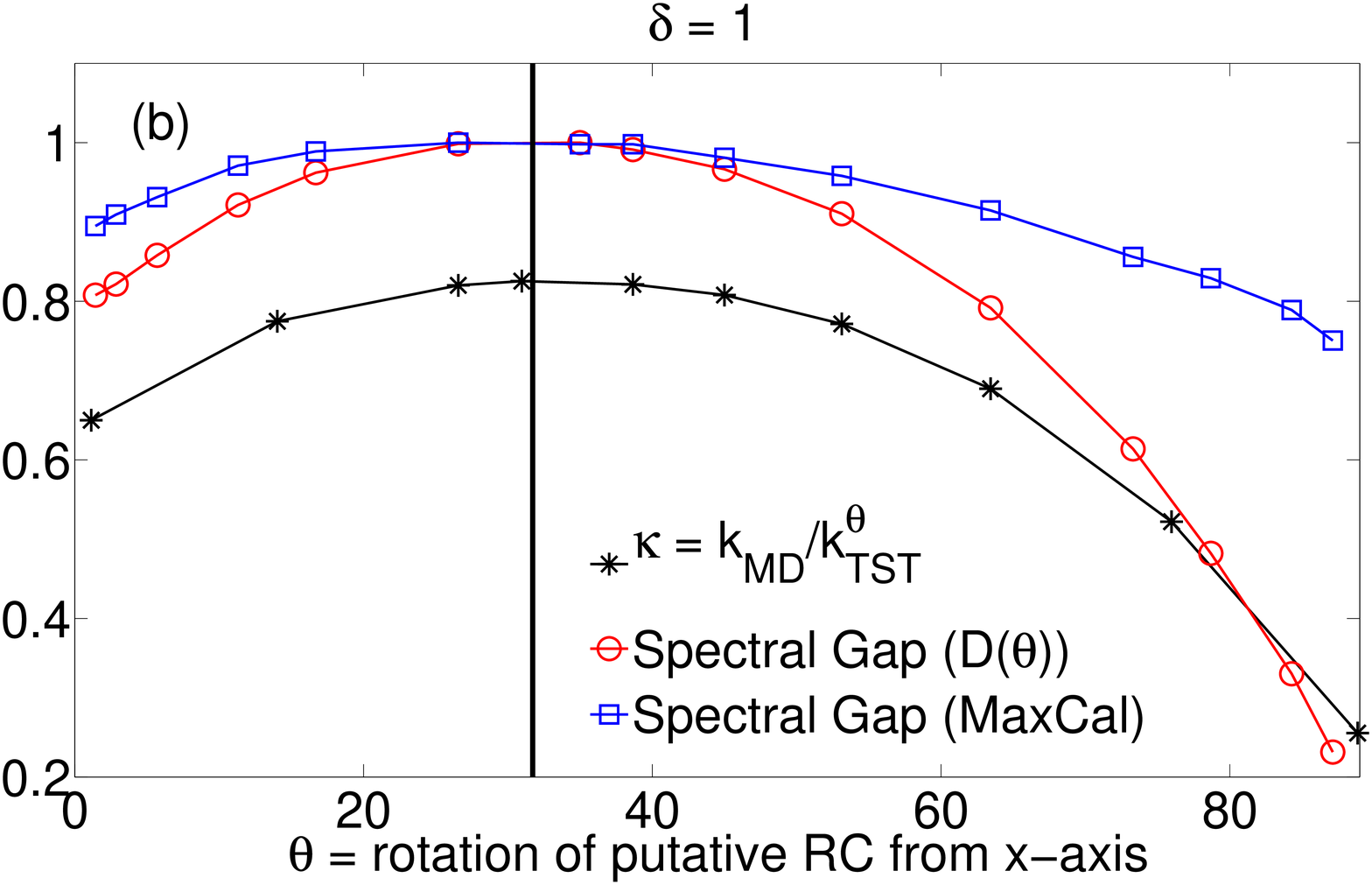} 
        \includegraphics[height=2.02in]{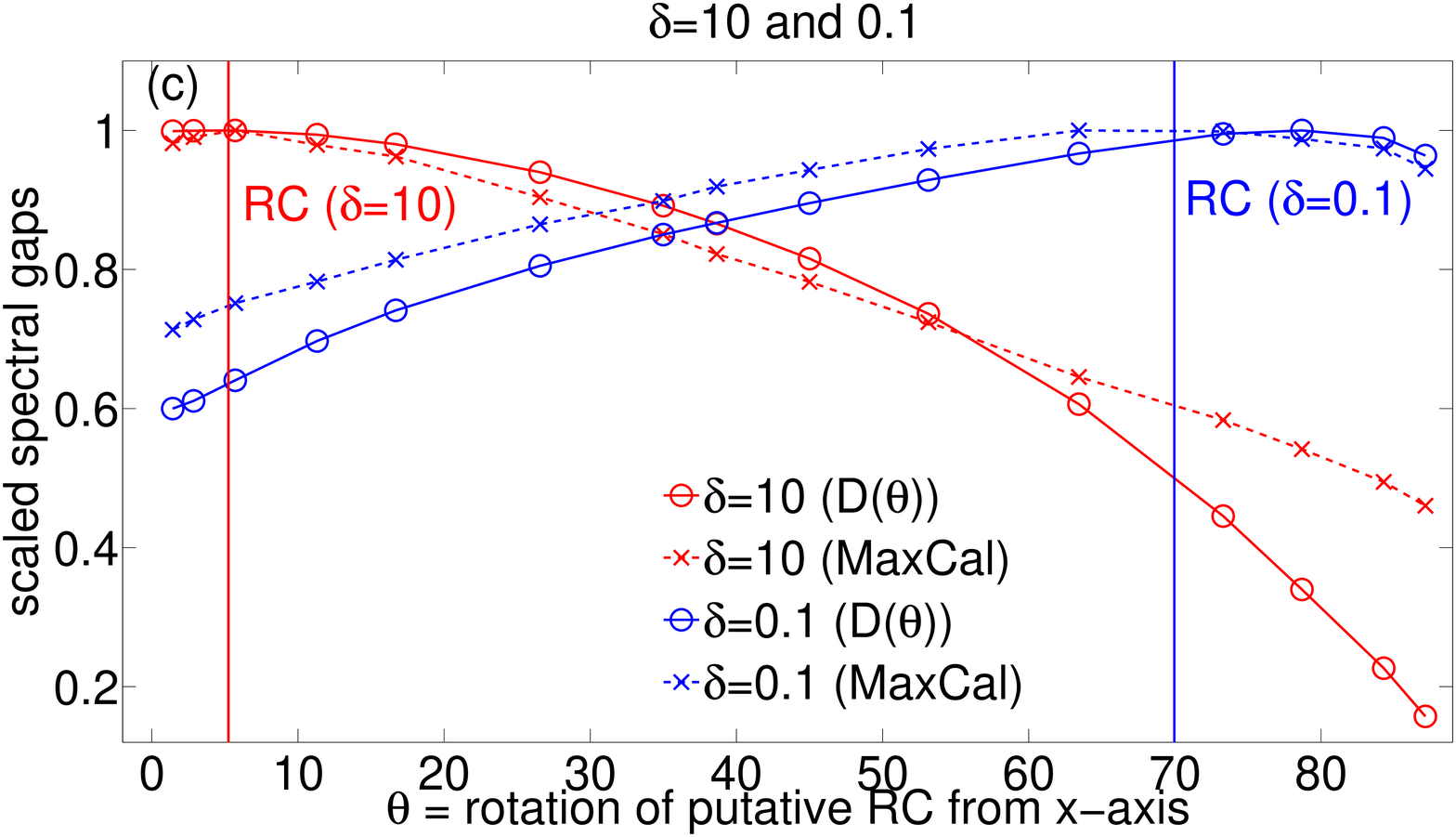} 
 	\includegraphics[height=2.02in]{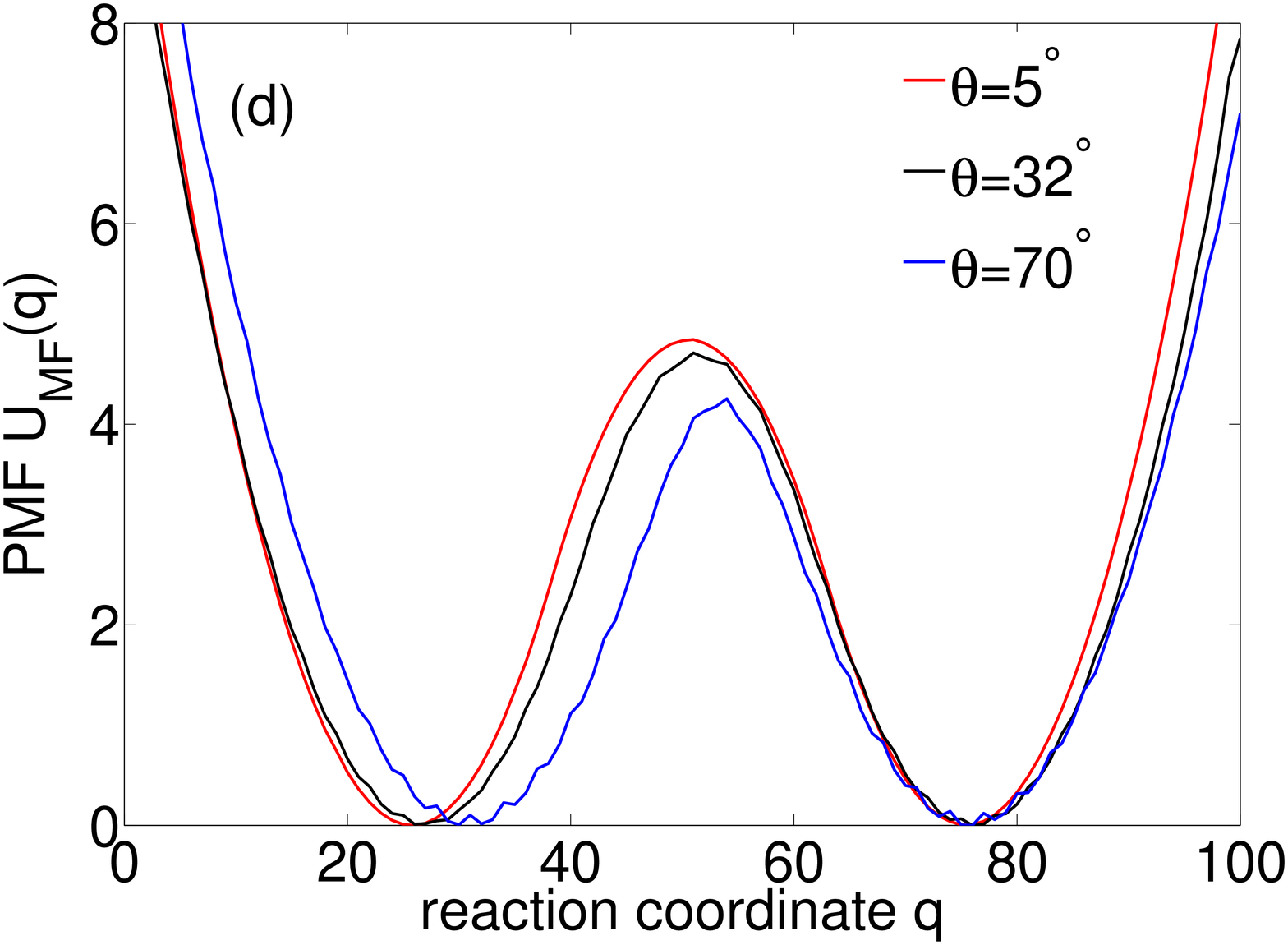} 
\caption{ For the Berezhkovskii-Szabo potential with various diffusion anisotropies $\delta =D{_y}/D{_x}$, SGOOP can accurately locate the true RC using Eq. \ref{rate_szabo} or Eq.\ref{rate_maxcal}. (a) shows the Berezhkovskii-Szabo potential $\beta U(x,y)$ from Ref. \cite{szabo_anisotropic}. Contours are drawn every 1 unit. The magenta colored dashed line denotes a putative RC $\hat{u}$, at an angle $\theta$ from the x-axis (solid black line). In (b) and (c), the vertical lines denote the benchmark calculations by Berezhkovskii-Szabo for various $\delta$ values. (b) gives various results for the isotropic case $\delta$ = 1.  The black stars are the transmission coefficients along various putative RCs $\hat{u}(\theta)$, given by the ratio of the rate constant $k_{MD}$ from long MD simulations, and the transition state theory (TST) rate constant $k_{TST}^{\theta}$ along $\hat{u}(\theta)$. (c) gives the spectral gaps for the anisotropic cases, with red and blue lines for $\delta$ = 10 and 1 respectively. The locations of the maximum spectral gaps are in excellent agreement with calculations of Berezhkovskii-Szabo marked with vertical lines. In (d), the respective 1-d PMFs along three putative RCs is provided.}
\label{fig:BS}
\end{figure*}

We finish this subsection by making the interesting observation that equations very similar to Eqs. \ref{rate_szabo} and \ref{rate_maxcal} have recently been used in the context of machine learning as well \cite{jascha_prl,jascha_picml}.

\subsection{Summarizing Spectral gap optimization of order parameters (SGOOP)}
\label{sgoop}
Given the full theoretical framework of the preceding sections, it is now easy to describe SGOOP \cite{sgoop}, which is a method to construct an optimal RC as a combination (linear or non-linear) of a set of many candidate order
parameters or collective variables $\Psi = (\Psi_1,\Psi_2,...,\Psi_d)$. This set is
assumed to be known \textit{a priori}. The working definition of the optimal RC that SGOOP follows was described in Secs. \ref{rc_def} and \ref{rc_dynamics}: (1) demarcate between  
various metastable states that the full system possesses, and
 (2) maximize the spectral gap of the eigenvalues of the associated master equation.
Once again, let $\{\lambda\}$ denote this set of eigenvalues, with  $\lambda_0=0<\lambda_1 \leq \lambda_2 \leq ...$. We refer to the quantity $e^{-\lambda_{n}}-e^{-\lambda_{n+1}}$ as the spectral gap,  where $n$ is the number of barriers apparent from the free energy estimate that are higher than a user-defined threshold (typically $\gtrsim k_B T$). This calculation is done without any pre-knowledge of the number of metastable states the system might have, and as such makes it easy to use SGOOP for systems with an arbitrary number of metastable states.

\begin{figure*}
  \centering
        \includegraphics[height=2.3in]{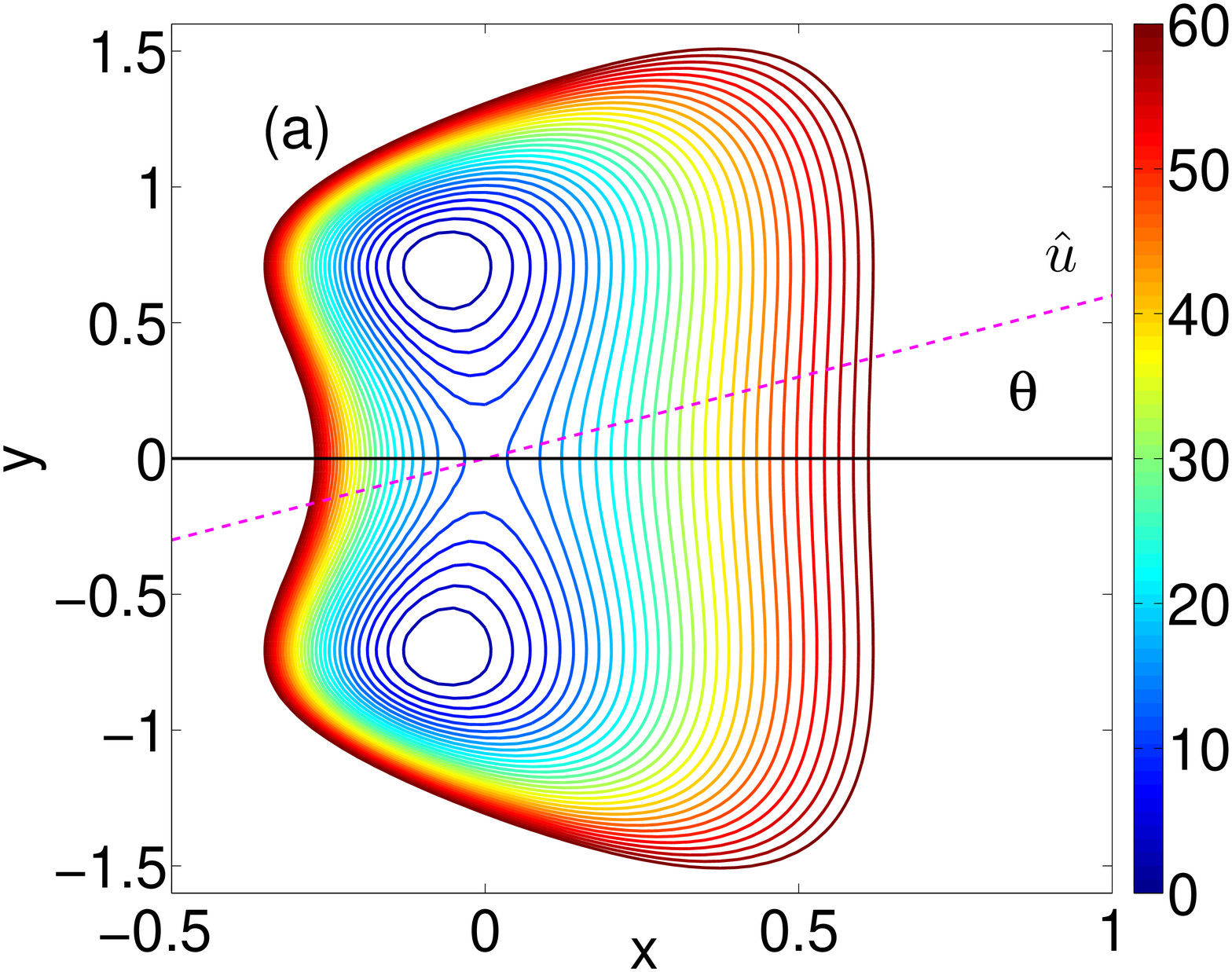} 
        \includegraphics[height=2.3in]{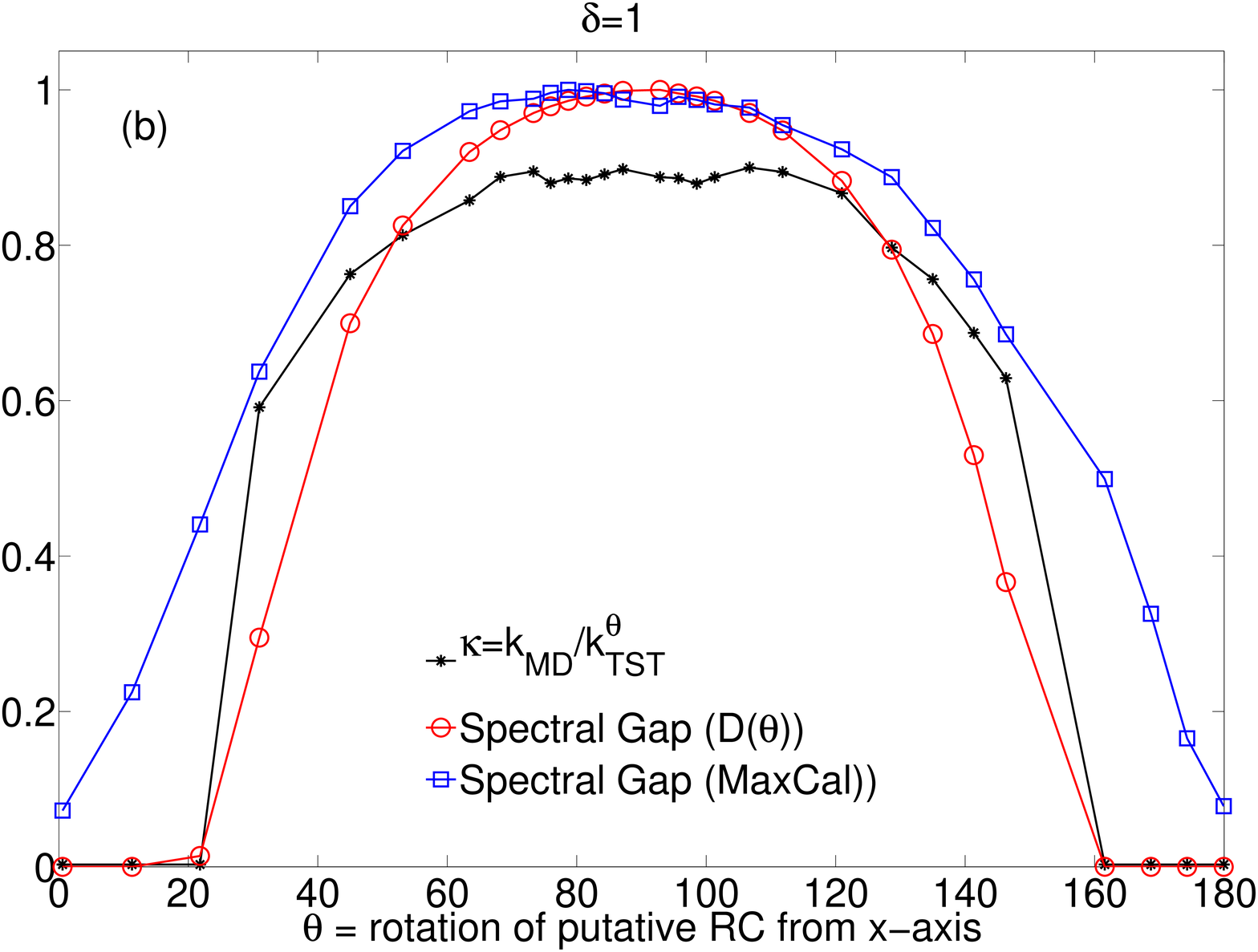} 
        \includegraphics[height=2.3in]{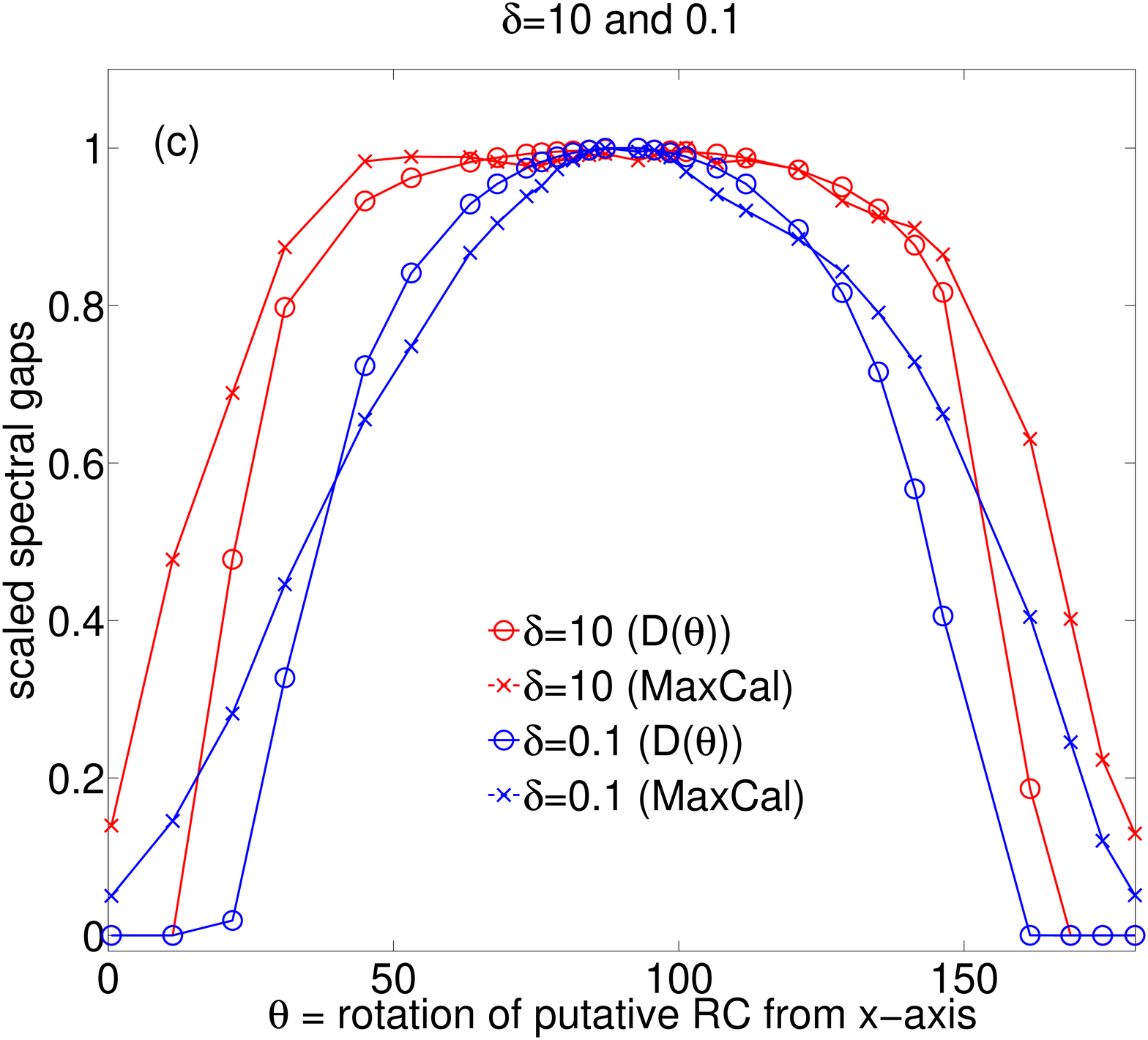} 
 	\includegraphics[height=2.3in]{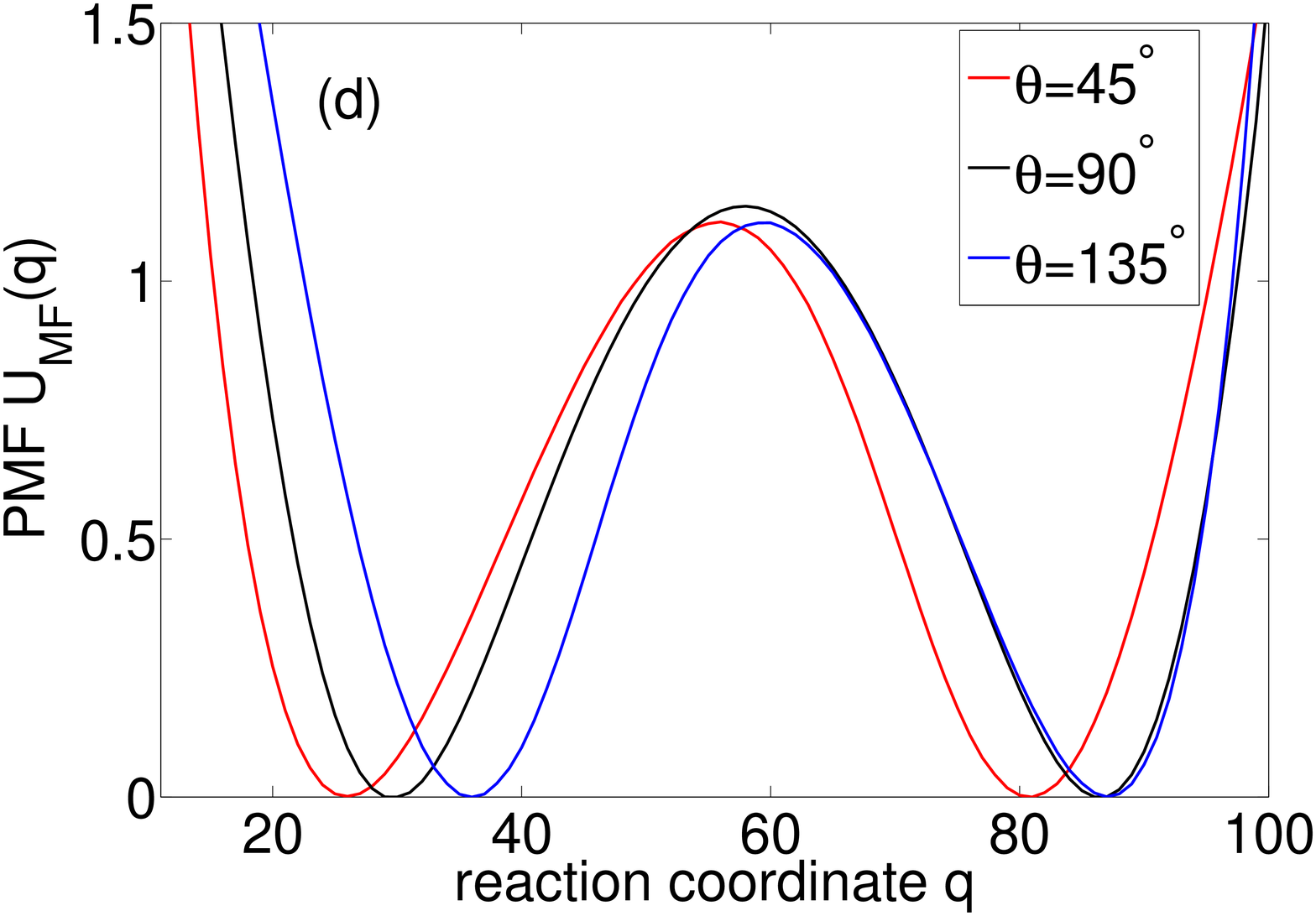} 
\caption{ For the Deleon-Berne potential with various diffusion anisotropies $\delta =D{_y}/D{_x}$, SGOOP can accurately locate the true RC using Eq. \ref{rate_szabo} or Eq.\ref{rate_maxcal}.  (a) shows the Deleon-Berne potential $\beta U(x,y)$ from Ref.\cite{kramers}. Contours are drawn every 1 unit. The magenta colored dashed line denotes a putative RC $\hat{u}$, at an angle $\theta$ from the x-axis (solid black line). (b) gives various results for the isotropic case $\delta$ = 1. The black stars are the transmission coefficients along various putative RCs $\hat{u}(\theta)$, given by the ratio of the rate constant $k_{MD}$ from long MD simulations, and the TST rate constant $k_{TST}^{\theta}$ along $\hat{u}(\theta)$. (c) gives the spectral gaps for the anisotropic cases, with red and blue lines for $\delta$ = 10 and 1 respectively. In (d), the respective 1-d PMFs along three putative RCs is provided.}
\label{fig:DB}
\end{figure*}

The first input to SGOOP is an estimate of the
stationary probability density (or equivalently the free energy) of
the system as a function of any putative RC $f(\Psi)$. For the model systems in this work 
$\Psi= \{x,y\}$ and the stationary density for any putative RC can be calculated analytically or numerically. For more complex systems \cite{sgoop,sgoop_fullerene}
SGOOP uses biased simulation\cite{tiwary_rewt}  performed along a
sub-optimal trial RC given by some linear or non-linear function
$f_0(\Psi)$, and then \textit{a posteriori} reweighting it along any putative RC $f(\Psi)$. The ability to perform computationally easy \textit {a posteriori} reweighting without repeating the simulations is crucial. The second input to SGOOP is some form of dynamical information. Depending on whether Eq. \ref{rate_szabo} or Eq. \ref{rate_maxcal} is used, this would respectively be the diffusivity tensor from Sec. \ref{rc_pmf}, or the average number of transitions $\langle N \rangle $
from Sec. \ref{szabo_maxcal}. Both these pieces of information are then used through either Eq. \ref{rate_szabo} or Eq. \ref{rate_maxcal} to obtain the eigenspectrum and hence the spectral gaps of the 
various trial CVs $f(\Psi)$. Through a post-processing optimization
procedure we then find the optimal RC as the $f(\Psi)$ which gives the
maximal spectral gap.

\referee{For the simple case with just one barrier in two dimensions, our method amounts to maximizing the difference between the first two nonzero eigenvalues of the master equation, which is in the spirt of other methods such as Ref. \cite{berezhkovskii2013diffusion}. In this work we implement SGOOP in two ways - by explicitly considering the diffusion tensor in the discretized Smoluchowski equation and through a Maximal Caliber framework, and demonstrate excellent agreement between both. The novelty in SGOOP partly comes from its ability to deal with more than one barrier which we show explicitly in the Results section, as well as through its use of the Maximal Caliber framework when the diffusion tensor is not known.}

\begin{figure*}
  \centering
        \includegraphics[height=2.11in]{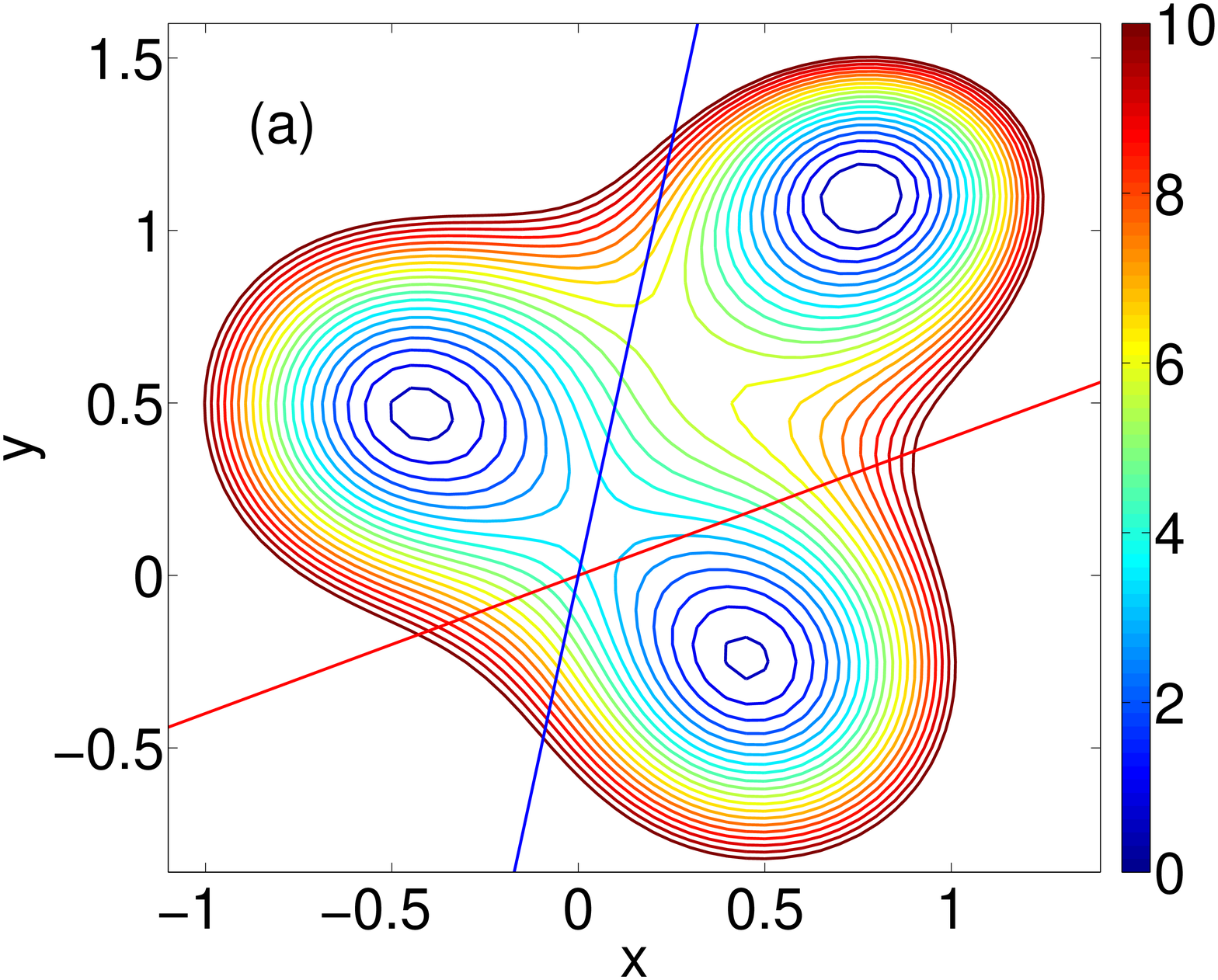} 
        \includegraphics[height=2.11in]{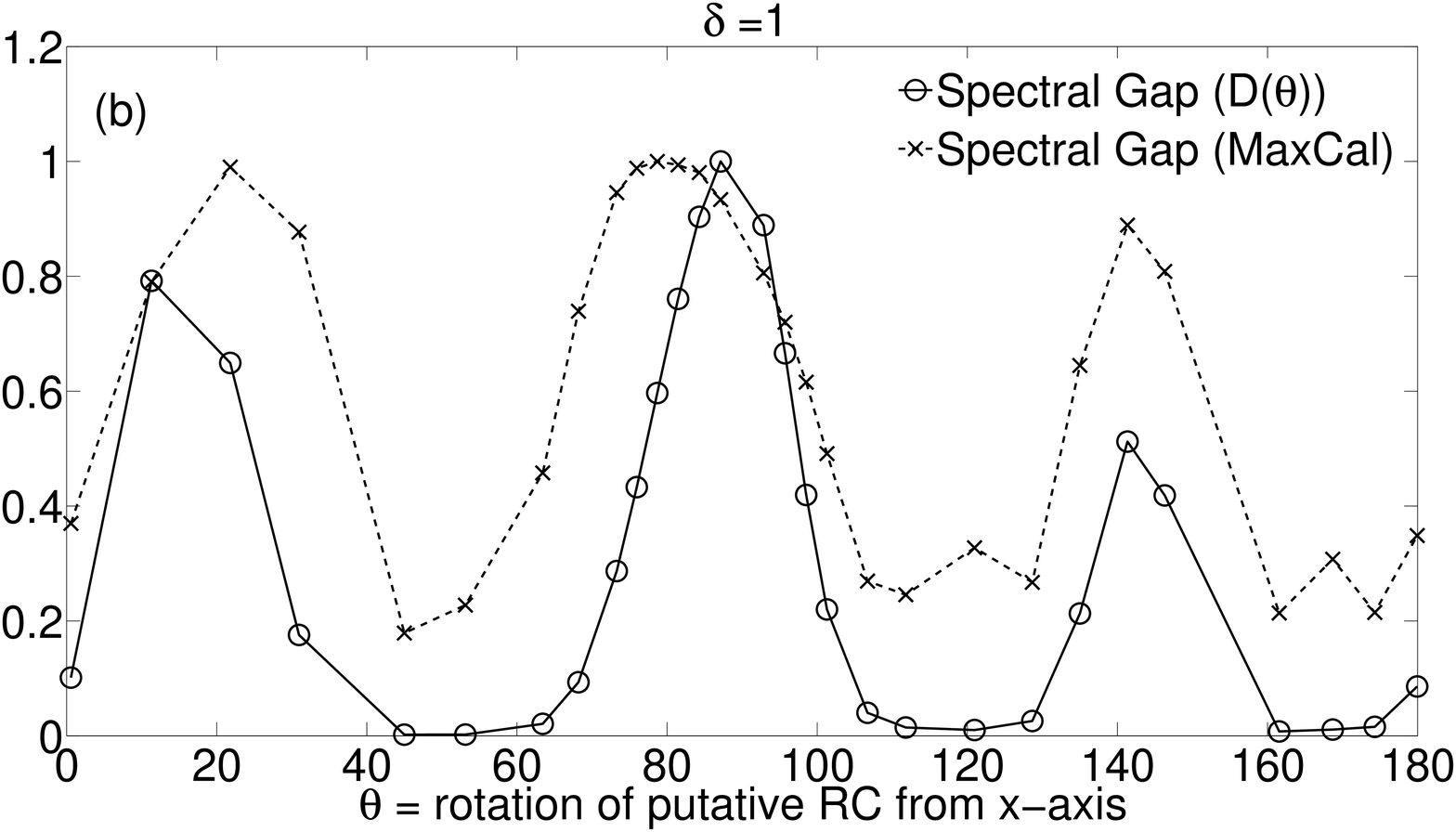} 
        \includegraphics[height=2.11in]{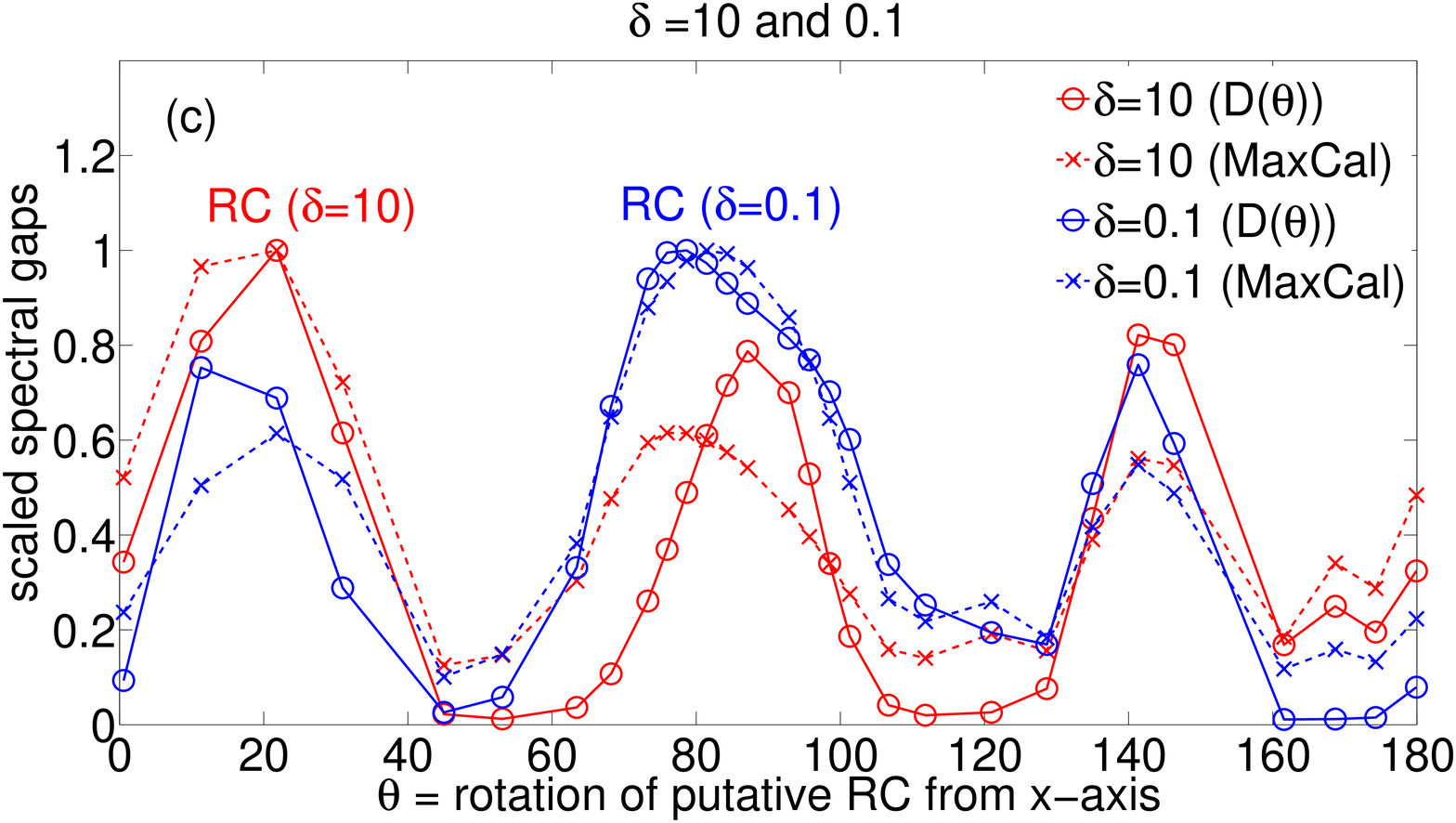} 
     \includegraphics[height=2.11in]{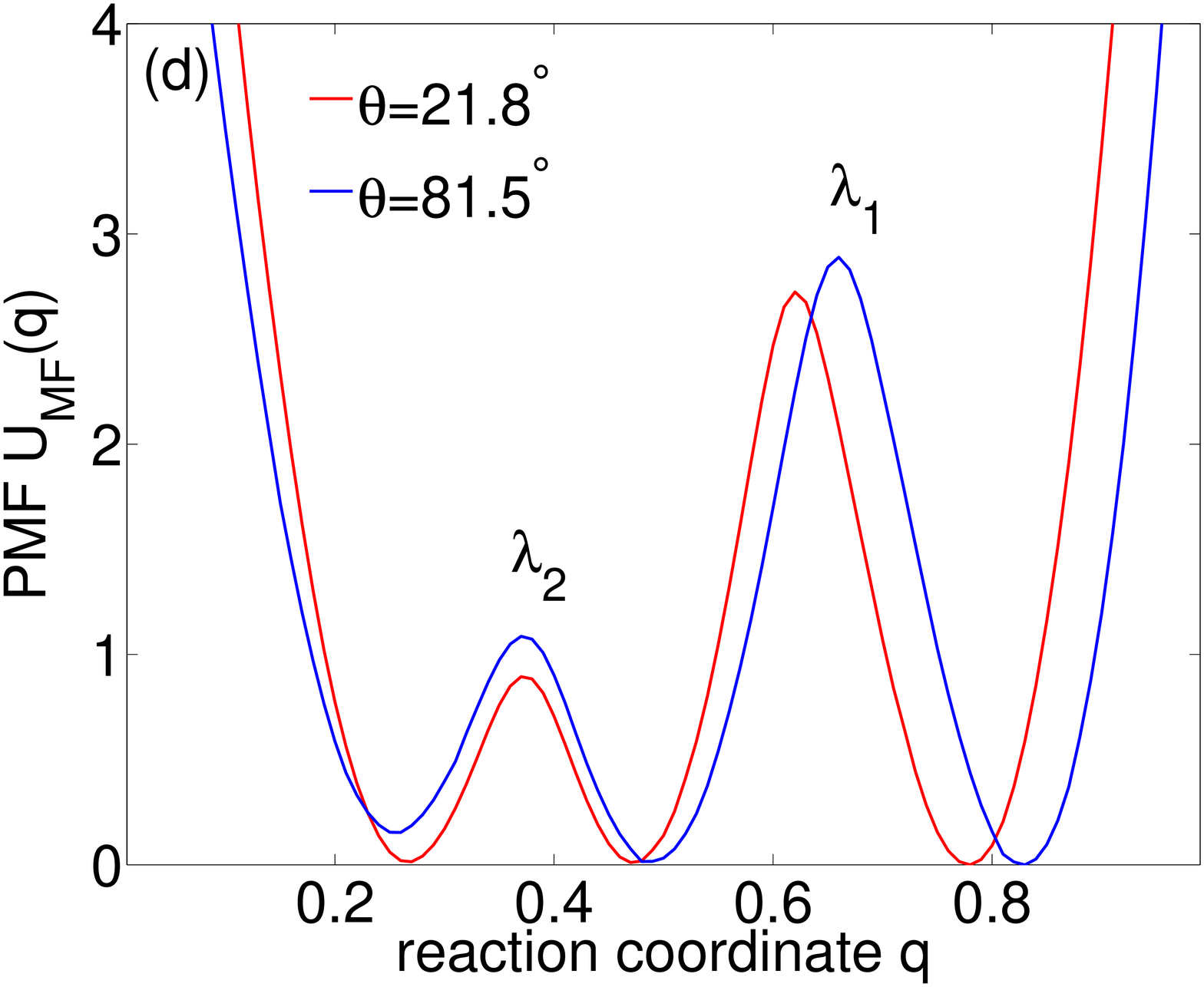} 
\caption{ For the 3-state potential with various diffusion anisotropies $\delta =D{_y}/D{_x}$, SGOOP can accurately locate the true RC using Eq. \ref{rate_szabo} or Eq.\ref{rate_maxcal}.  (a) shows the 3-state potential $\beta U(x,y)$. Contours are drawn every 0.5 unit. Two putative RCs $\hat{u}$ are shown, at an angles 21.8$^{\circ}$ and 81.5$^{\circ}$ from the x-axis. (b) gives various results for the isotropic case $\delta$ = 1. (c) gives the spectral gaps for the anisotropic cases, with red and blue lines for $\delta$ = 10 and 1 respectively. The RC for these two anisotropies is found to be at $\theta$ = 21.8$^{\circ}$ and 81.5$^{\circ}$ (marked in (a)). In (d), the respective 1-d PMFs along two putative RCs is provided.}
\label{fig:3state}
\end{figure*}

\section{Results}
\label{results}
Previous applications of SGOOP have been limited to considering only the stationary density information and ignoring dynamical effects. In this work we consider three model 2-dimensional potentials undergoing Langevin dynamics coupled to a bath, with distinct dynamical effects due to diffusion anisotropy, and demonstrate how SGOOP can be used to accurately predict the effects of diffusion anisotropy on the RC. The first two are 2-state potentials, namely the Berezhkovskii-Szabo (BS) potential \cite{szabo_anisotropic} and the Deleon-Berne (DB) potential \cite{deleonberne}. The third potential is a 3-state potential which we describe later. The set of candidate order parameters is given by $\Psi= \{x,y\}$, and for each potential, we identify the optimal RC as a linear combination $f(x,y)= ax+by$ for different diffusion anisotropies $\delta = {D_y / D_x}$. The This amounts to calculating the optimal angle $\theta$ from $\hat{x}$ for RC given by $\hat{u} = \hat{x} cos(\theta)+\hat{y} sin(\theta)$ for $\delta=$ 0.1, 1 and 10. For all the potentials, Newton's laws of motion were integrated per Langevin dynamics\cite{bussi_langevin} with a timestep 0.01 unit and different values of anisotropic friction coefficients $\gamma_x,\gamma_y$. Since $\delta = {D_y / D_x}$, we have the relation $\delta = {\gamma_x / \gamma_y}$. Corresponding to $\delta=$ 0.1, 1 and 10, we took ($\gamma_x,\gamma_y$) = (2,20), (5,5) and (20,2) respectively in inverse time units. All of these simulations correspond to the overdamped friction regime \cite{kramers}.

We consider two different 2-state potentials because of the structure of the diffusion tensor. In one of them (DB), the two directions $x$ and $y$ are parallel/perpendicular to the line through the saddle connecting the two minima, while in the other (BS), both the directions are  at a tilt with respect to this line. This simple difference, as we show later, leads to an arguably profound difference in how the RC depends on diffusion anisotropy. For the 2-state potentials we also perform explicit calculations of the transmission coefficient given by the ratio of the rate constant $k_{MD}$ from long MD simulations, and the TST rate constant $k_{TST}^{\theta}$ along various $\hat{u}(\theta)$.

\subsection{Berezhkovskii-Szabo (BS) potential}
The first potential we consider is a 2-state potential (Fig. \ref{fig:BS}(a)) introduced by Berezhkovskii and Szabo \cite{szabo_anisotropic}. We start with this potential because the authors in Ref. \onlinecite{szabo_anisotropic} have explicitly calculated the dependence of the RC on the anisotropy parameter $\delta$ through a different method. Thus it serves as an excellent benchmark for our work. In Fig. \ref{fig:BS}(a) the magenta colored dashed line denotes a putative RC $\hat{u} = \hat{x} cos(\theta)+\hat{y} sin(\theta)$, at an angle $\theta$ from the x-axis (solid black line). The potential is detailed in Ref. \onlinecite{szabo_anisotropic}. Here we perform Langevin dynamics at inverse temperature $\beta=1$.

In Fig. \ref{fig:BS} we give the detailed results for this potential. Fig. \ref{fig:BS}(b) shows the results for the isotropic diffusion case. Here the use of SGOOP with Eq. \ref{rate_szabo} or  Eq. \ref{rate_maxcal} gives the same optimal RC ($\theta$=32$^\circ$) in very good agreement with the benchmark calculation in Ref. \cite{szabo_anisotropic}. We also provide here our calculations of the transmission coefficients along various putative RCs $\hat{u}(\theta)$, given by the ratio of the rate constant $k_{MD}$, and the TST rate constant $k_{TST}^{\theta}$ along $\hat{u}(\theta)$. It is very interesting that the scaled spectral gap through SGOOP follows roughly the dependence of the transmission coefficient on $\theta$.
Fig. \ref{fig:BS}(c) gives the same information as (b) but for the anisotropic cases. Here as well the agreement with the benchmark calculation in Ref. \cite{szabo_anisotropic} is excellent, with the optimal RC at $\theta$=5$^\circ$  and $\theta$=70$^\circ$ respectively for $\delta$=10 and 0.1 respectively. Naturally, the transmission coefficient ($\kappa = k_{MD} / k_{TST}^{\theta}$) would have the same qualitative behavior including location of maximal value at $\theta$=32$^\circ$, irrespective of the extent of anisotropy.  This is because $k_{TST}^{\theta}$ does not depend on the anisotropy in $D$ but only on 
the static details through the PMF for the putative RC corresponding to $\theta$, and $k_{MD}$ is a measurement independent of the choice of RC i.e. $\theta$. This leads to the possibly obvious but important observation that while maximizing the value of the transmission coefficient still gives the best TST description (i.e. variational transition state theory), it might not give the best reaction coordinate as it does not take into account dynamical effects through the anisotropy of the diffusion coefficients. Finally, in Fig. \ref{fig:BS} (d), we provide the respective 1-d PMFs along three putative RCs. It is interesting to see here how the effect of a smaller barrier - which would be ideal for a 2-state problem from the purpose of optimizing transmission coefficient - is compensated by the slower diffusion along a certain direction.

\subsection{Deleon-Berne (DB) potential}
The second potential we consider is a 2-state potential (Fig. \ref{fig:DB}(a)) introduced by Deleon and Berne \cite{deleonberne}.  The potential is detailed in Ref. \onlinecite{kramers}. Here we perform Langevin dynamics at inverse temperature $\beta=5$. In Fig. \ref{fig:DB} we give the detailed results for this potential. Fig. \ref{fig:DB}(b) shows the results for the isotropic diffusion case. Here the use of SGOOP with Eq. \ref{rate_szabo} or  Eq. \ref{rate_maxcal} gives the same optimal RC ($\theta$=$90^\circ$) in very good agreement with the transmission coefficient calculation. It is again very interesting that the scaled spectral gap through SGOOP follows roughly the dependence of the transmission coefficient on $\theta$. Fig. \ref{fig:DB}(c) gives the same information as (b) for the anisotropic cases. 

For this potential, unlike the Berezhkovskii-Szabo potential, changing the diffusion anisotropy does not change the precise location of the optimal RC. However, it does change the tolerance in quality of RC one has in terms of deviating from the optimal location. Thus as can be seen in Fig. \ref{fig:DB}(d),  the spectral gap versus $\theta$ profiles for the $\delta$=10 and 0.1 cases are broader and narrower respectively than the $\delta$=1 profile in Fig. \ref{fig:DB}(b). Thus when the diffusion along $y$ is slower ($\delta=$0.1), the penalty for deviating from the optimal RC  $\theta$ = $90^{\circ}$  is much higher, and correspondingly it is much lower in the case when diffusion along $x$ is slower ($\delta=$10). We further discuss this point in Sec. \ref{discussion}. In Fig. \ref{fig:DB} (d), we provide the respective 1-d PMFs along three putative RCs. As can be seen here, the RC with $\theta$ = 90$^{\circ}$ gives the highest barrier and thus corresponds to the optimal RC for the isotropic diffusion case. 

One interesting observation that can be made from the two 2-state problems we looked at pertains to the qualitative difference in their response to the diffusion anisotropy. The  Berezhkovskii-Szabo (BS) potential showed a very clear dependence of the optimal RC on extent of diffusion anisotropy, while the Deleon-Berne (DB) potential had the same optimal RC location irrespective of the extent of anisotropy. Apart from the difference in the precise functional form, both of these are simple 2-state potentials, fairly similar to each other except that in the BS potential, the diffusion tensor non-zero components $x$ and $y$ are not aligned along the normal mode of the PES. In the DB potential, one of the diffusion tensor non-zero components ($y$) is aligned along the normal mode of the PES. 

\subsection{3-state potential}
The final potential we consider is a 3-state potential given by the following functional form:
\begin{align}
U(x,y) = &-16(e^{-2(x+.5)^2-2(y-.5)^2}) \nonumber \\ 
 &-18(e^{-2(x-.8)^2-2(y-1.2)^2}) \nonumber \\ 
 &-16(e^{-2(x-.5)^2-2(y+.3)^2})  +   0.5(x^6 + y^6)
\label{3state}
\end{align}

\begin{figure}
  \centering
        \includegraphics[height=3.2in]{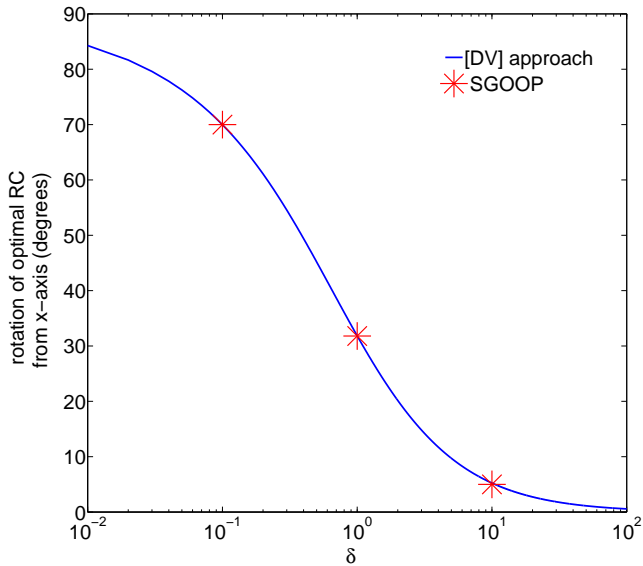} 
\caption{Blue line shows the rotation of the optimal RC from $x-$axis for the BS potential, as obtained by diagonalizing the $DV$ matrix, as described in Sec. \ref{additional}. Red asterisks mark estimates of the optimal RC from SGOOP.}
\label{fig:DV}
\end{figure}

 Here we perform Langevin dynamics at inverse temperature $\beta=1.25$. In Fig. \ref{fig:3state} we give the detailed results for this potential. Fig. \ref{fig:3state}(b) shows the results for the isotropic diffusion case. Here the use of SGOOP with Eq. \ref{rate_szabo} or  Eq. \ref{rate_maxcal} gives the same optimal RC ($\theta$=81.5$^\circ$). 

Note that there are two other peaks in this profile, signifying the presence of multiple barriers and thus different families of possible RC. These peaks are however not as dominant as the optimal RC. Fig. \ref{fig:3state}(c) gives the same information as (b) for the anisotropic cases. For this potential, when we change the diffusion anisotropy to $\delta=$10 (i.e. diffusion along $x$ is slower), we see a flip in the relative strengths of the peaks at $\theta$=81.5$^\circ$ and 
$\theta$=21.8$^\circ$, and the optimal RC switches to $\theta$=21.8$^\circ$. For the opposite end when $\delta=$0.1 (i.e. diffusion along $y$ is slower), the optimal RC is same as in the isotropic case. The same results are found through the use of Eq. \ref{rate_szabo} or Eq. \ref{rate_maxcal}.

In Fig. \ref{fig:3state} (d), we provide the respective 1-d PMFs along the two putative RCs obtained for $\theta$=81.5$^\circ$ (blue line) and $\theta$=21.8$^\circ$ (red line). In contrast to the other two potentials, now we have two barriers. Similar interpretations can be made here as for the 2-state potentials. The RC for $\delta$=1 has higher barriers (corresponding to both the eigenvalues $\lambda_1$ and $\lambda_2$). However for the $\delta$=10 case it is the effect of having slower diffusion along $x$ that leads to the putative RC with lower free energy barriers becoming the optimal RC.

\subsection{Additional treatment of the Berezhkovskii-Szabo (BS) potential}
\label{additional}
In this subsection, we revisit the BS potential from yet another perspective which has been used by workers including Hynes, Szabo and others \cite{van1982reactive,szabo_anisotropic}, and has been described in those papers. Unlike SGOOP, this treatment is applicable only if (1) the reactants and products are separated by a single saddle point, and (2) the dynamics is in the high friction limit. The key idea here is to consider the matrix $DV$ where $D$ is the full diffusion tensor, and $V$ is matrix of second derivatives of the multi-dimensional potential at its saddle point, with det $V<$  0. Both these matrices could in general have off-diagonal cross terms. By then solving for the only negative eigenvalue and associate eigenvector of the operator $DV$, we get the reaction coordinate for the given potential. We do this exercise for various diffusion anisotropies as quantified by the value of the parameter $\delta$ in the diffusion tensor \[ \left( \begin{array}{cc}
D_x & 0\\
0 & \delta D_x \end{array} \right)\] 
In Fig. \ref{fig:DV} we provide the rotation of the optimal RC from $x-$axis for the BS potential as obtained by diagonalizing the $DV$ matrix and calculating the orientation of the eigenvector with the only negative eigenvalue. In this figure we have also overlaid the estimate of the optimal RC from SGOOP and the two estimates are in excellent agreement.

\section{Discussion}
\label{discussion}
In this work we have demonstrated how using our recent method SGOOP \cite{sgoop} we can reliably and accurately calculate the effect of diffusion anisotropy on the reaction coordinate for different model potential energy landscapes with high and arbitrarily many energy barriers, and arbitrary diffusion anisotropies. Where possible, we validated the SGOOP-predicted RC with published results using different methods for RC calculation\cite{szabo_anisotropic}, and also with extensive calculations of the transmission coefficients along different putative RCs. We showed how the dynamic information in SGOOP can be explicitly implemented either through explicit knowledge of the diffusion tensor, or through a Maximum Caliber framework\cite{caliber1,dixit2015inferring}. We found excellent agreement between the results for all three potentials using both these methods. Unlike other methods for optimizing RC that we are aware of, SGOOP was also shown to give an approximate quantification of how tolerant is the system with respect to deviations from the optimal RC. 

Clearly, knowing the optimal RC is important from the perspective of kinetics calculations. However we believe that diffusion anisotropy and how it is manifested in the interplay between statics and dynamics might be an overlooked but vital problem in free energy sampling as well (barring few notable exceptions such as Ref. \cite{poon2015accelerated}), where one is often faced with a choice of many qualitatively different order parameters or collective variables \cite{arpc_meta,tiwary2016review,mccarty2017variational}. Barring very specialized conditions, these variables will naturally have different diffusivities. In such situations, a method like SGOOP should be very useful for accurate and efficient free energy sampling. A free energy profile sampled along the optimal RC will be easier to compute through umbrella sampling or metadynamics, and features in this profile will be of more direct use as well for drawing physical conclusions. This will be the subject of future investigations probing the thermodynamics and kinetics of complex systems.

\vspace{.2in}

\textbf{ACKNOWLEDGMENTS}\\

This work was supported by grants from the National Institutes of
Health [NIH-GM4330] and the Extreme Science and Engineering Discovery
Environment (XSEDE) [TG-MCA08X002]. The authors thanks Purushottam Dixit for helpful discussions.


\begin{thebibliography}{39}%
\makeatletter
\providecommand \@ifxundefined [1]{%
 \@ifx{#1\undefined}
}%
\providecommand \@ifnum [1]{%
 \ifnum #1\expandafter \@firstoftwo
 \else \expandafter \@secondoftwo
 \fi
}%
\providecommand \@ifx [1]{%
 \ifx #1\expandafter \@firstoftwo
 \else \expandafter \@secondoftwo
 \fi
}%
\providecommand \natexlab [1]{#1}%
\providecommand \enquote  [1]{``#1''}%
\providecommand \bibnamefont  [1]{#1}%
\providecommand \bibfnamefont [1]{#1}%
\providecommand \citenamefont [1]{#1}%
\providecommand \href@noop [0]{\@secondoftwo}%
\providecommand \href [0]{\begingroup \@sanitize@url \@href}%
\providecommand \@href[1]{\@@startlink{#1}\@@href}%
\providecommand \@@href[1]{\endgroup#1\@@endlink}%
\providecommand \@sanitize@url [0]{\catcode `\\12\catcode `\$12\catcode
  `\&12\catcode `\#12\catcode `\^12\catcode `\_12\catcode `\%12\relax}%
\providecommand \@@startlink[1]{}%
\providecommand \@@endlink[0]{}%
\providecommand \url  [0]{\begingroup\@sanitize@url \@url }%
\providecommand \@url [1]{\endgroup\@href {#1}{\urlprefix }}%
\providecommand \urlprefix  [0]{URL }%
\providecommand \Eprint [0]{\href }%
\providecommand \doibase [0]{http://dx.doi.org/}%
\providecommand \selectlanguage [0]{\@gobble}%
\providecommand \bibinfo  [0]{\@secondoftwo}%
\providecommand \bibfield  [0]{\@secondoftwo}%
\providecommand \translation [1]{[#1]}%
\providecommand \BibitemOpen [0]{}%
\providecommand \bibitemStop [0]{}%
\providecommand \bibitemNoStop [0]{.\EOS\space}%
\providecommand \EOS [0]{\spacefactor3000\relax}%
\providecommand \BibitemShut  [1]{\csname bibitem#1\endcsname}%
\let\auto@bib@innerbib\@empty
\bibitem [{\citenamefont {Henkelman}, \citenamefont {Uberuaga},\ and\
  \citenamefont {J{\'o}nsson}(2000)}]{neb}%
  \BibitemOpen
  \bibfield  {author} {\bibinfo {author} {\bibfnamefont {G.}~\bibnamefont
  {Henkelman}}, \bibinfo {author} {\bibfnamefont {B.~P.}\ \bibnamefont
  {Uberuaga}}, \ and\ \bibinfo {author} {\bibfnamefont {H.}~\bibnamefont
  {J{\'o}nsson}},\ }\href@noop {} {\bibfield  {journal} {\bibinfo  {journal}
  {J. Chem. Phys.}\ }\textbf {\bibinfo {volume} {113}},\ \bibinfo {pages}
  {9901} (\bibinfo {year} {2000})}\BibitemShut {NoStop}%
\bibitem [{\citenamefont {Branduardi}, \citenamefont {Gervasio},\ and\
  \citenamefont {Parrinello}(2007)}]{fromatob}%
  \BibitemOpen
  \bibfield  {author} {\bibinfo {author} {\bibfnamefont {D.}~\bibnamefont
  {Branduardi}}, \bibinfo {author} {\bibfnamefont {F.~L.}\ \bibnamefont
  {Gervasio}}, \ and\ \bibinfo {author} {\bibfnamefont {M.}~\bibnamefont
  {Parrinello}},\ }\href@noop {} {\bibfield  {journal} {\bibinfo  {journal} {J
  Chem Phys}\ }\textbf {\bibinfo {volume} {126}},\ \bibinfo {pages} {054103}
  (\bibinfo {year} {2007})}\BibitemShut {NoStop}%
\bibitem [{\citenamefont {Weinan}, \citenamefont {Ren},\ and\ \citenamefont
  {Vanden-Eijnden}(2007)}]{weinan2007}%
  \BibitemOpen
  \bibfield  {author} {\bibinfo {author} {\bibfnamefont {E.}~\bibnamefont
  {Weinan}}, \bibinfo {author} {\bibfnamefont {W.}~\bibnamefont {Ren}}, \ and\
  \bibinfo {author} {\bibfnamefont {E.}~\bibnamefont {Vanden-Eijnden}},\
  }\href@noop {} {\bibfield  {journal} {\bibinfo  {journal} {The Journal of
  Chemical Physics}\ } (\bibinfo {year} {2007})}\BibitemShut {NoStop}%
\bibitem [{\citenamefont {van~der Zwan}\ and\ \citenamefont
  {Hynes}(1982)}]{van1982reactive}%
  \BibitemOpen
  \bibfield  {author} {\bibinfo {author} {\bibfnamefont {G.}~\bibnamefont
  {van~der Zwan}}\ and\ \bibinfo {author} {\bibfnamefont {J.~T.}\ \bibnamefont
  {Hynes}},\ }\href@noop {} {\bibfield  {journal} {\bibinfo  {journal} {The
  Journal of Chemical Physics}\ }\textbf {\bibinfo {volume} {77}},\ \bibinfo
  {pages} {1295} (\bibinfo {year} {1982})}\BibitemShut {NoStop}%
\bibitem [{\citenamefont {Berezhkovskii}\ and\ \citenamefont
  {Szabo}(2005)}]{szabo_anisotropic}%
  \BibitemOpen
  \bibfield  {author} {\bibinfo {author} {\bibfnamefont {A.}~\bibnamefont
  {Berezhkovskii}}\ and\ \bibinfo {author} {\bibfnamefont {A.}~\bibnamefont
  {Szabo}},\ }\href@noop {} {\bibfield  {journal} {\bibinfo  {journal} {The
  Journal of chemical physics}\ }\textbf {\bibinfo {volume} {122}},\ \bibinfo
  {pages} {014503} (\bibinfo {year} {2005})}\BibitemShut {NoStop}%
\bibitem [{\citenamefont {Hynes}(1985)}]{hynes1985chemical}%
  \BibitemOpen
  \bibfield  {author} {\bibinfo {author} {\bibfnamefont {J.~T.}\ \bibnamefont
  {Hynes}},\ }\href@noop {} {\bibfield  {journal} {\bibinfo  {journal} {Annual
  Review of Physical Chemistry}\ }\textbf {\bibinfo {volume} {36}},\ \bibinfo
  {pages} {573} (\bibinfo {year} {1985})}\BibitemShut {NoStop}%
\bibitem [{\citenamefont {Kl/osek-Dygas}\ \emph {et~al.}(1989)\citenamefont
  {Kl/osek-Dygas}, \citenamefont {Hoffman}, \citenamefont {Matkowsky},
  \citenamefont {Nitzan}, \citenamefont {Ratner},\ and\ \citenamefont
  {Schuss}}]{kl1989diffusion}%
  \BibitemOpen
  \bibfield  {author} {\bibinfo {author} {\bibfnamefont {M.}~\bibnamefont
  {Kl/osek-Dygas}}, \bibinfo {author} {\bibfnamefont {B.}~\bibnamefont
  {Hoffman}}, \bibinfo {author} {\bibfnamefont {B.}~\bibnamefont {Matkowsky}},
  \bibinfo {author} {\bibfnamefont {A.}~\bibnamefont {Nitzan}}, \bibinfo
  {author} {\bibfnamefont {M.~A.}\ \bibnamefont {Ratner}}, \ and\ \bibinfo
  {author} {\bibfnamefont {Z.}~\bibnamefont {Schuss}},\ }\href@noop {}
  {\bibfield  {journal} {\bibinfo  {journal} {The Journal of chemical physics}\
  }\textbf {\bibinfo {volume} {90}},\ \bibinfo {pages} {1141} (\bibinfo {year}
  {1989})}\BibitemShut {NoStop}%
\bibitem [{\citenamefont {Johnson}\ and\ \citenamefont
  {Hummer}(2012)}]{johnson2012characterization}%
  \BibitemOpen
  \bibfield  {author} {\bibinfo {author} {\bibfnamefont {M.~E.}\ \bibnamefont
  {Johnson}}\ and\ \bibinfo {author} {\bibfnamefont {G.}~\bibnamefont
  {Hummer}},\ }\href@noop {} {\bibfield  {journal} {\bibinfo  {journal} {The
  Journal of Physical Chemistry B}\ }\textbf {\bibinfo {volume} {116}},\
  \bibinfo {pages} {8573} (\bibinfo {year} {2012})}\BibitemShut {NoStop}%
\bibitem [{\citenamefont {Peters}\ \emph {et~al.}(2013)\citenamefont {Peters},
  \citenamefont {Bolhuis}, \citenamefont {Mullen},\ and\ \citenamefont
  {Shea}}]{peters2013reaction}%
  \BibitemOpen
  \bibfield  {author} {\bibinfo {author} {\bibfnamefont {B.}~\bibnamefont
  {Peters}}, \bibinfo {author} {\bibfnamefont {P.~G.}\ \bibnamefont {Bolhuis}},
  \bibinfo {author} {\bibfnamefont {R.~G.}\ \bibnamefont {Mullen}}, \ and\
  \bibinfo {author} {\bibfnamefont {J.-E.}\ \bibnamefont {Shea}},\ }\href@noop
  {} {\bibfield  {journal} {\bibinfo  {journal} {The Journal of chemical
  physics}\ }\textbf {\bibinfo {volume} {138}},\ \bibinfo {pages} {054106}
  (\bibinfo {year} {2013})}\BibitemShut {NoStop}%
\bibitem [{\citenamefont {Echeverria}, \citenamefont {Makarov},\ and\
  \citenamefont {Papoian}(2014)}]{echeverria2014concerted}%
  \BibitemOpen
  \bibfield  {author} {\bibinfo {author} {\bibfnamefont {I.}~\bibnamefont
  {Echeverria}}, \bibinfo {author} {\bibfnamefont {D.~E.}\ \bibnamefont
  {Makarov}}, \ and\ \bibinfo {author} {\bibfnamefont {G.~A.}\ \bibnamefont
  {Papoian}},\ }\href@noop {} {\bibfield  {journal} {\bibinfo  {journal}
  {Journal of the American Chemical Society}\ }\textbf {\bibinfo {volume}
  {136}},\ \bibinfo {pages} {8708} (\bibinfo {year} {2014})}\BibitemShut
  {NoStop}%
\bibitem [{\citenamefont {Northrup}\ and\ \citenamefont
  {McCammon}(1983)}]{northrup1983saddle}%
  \BibitemOpen
  \bibfield  {author} {\bibinfo {author} {\bibfnamefont {S.~H.}\ \bibnamefont
  {Northrup}}\ and\ \bibinfo {author} {\bibfnamefont {J.~A.}\ \bibnamefont
  {McCammon}},\ }\href@noop {} {\bibfield  {journal} {\bibinfo  {journal} {The
  Journal of Chemical Physics}\ }\textbf {\bibinfo {volume} {78}},\ \bibinfo
  {pages} {987} (\bibinfo {year} {1983})}\BibitemShut {NoStop}%
\bibitem [{\citenamefont {Berezhkovskii}, \citenamefont {Berezhkovskii},\ and\
  \citenamefont {Zitzerman}(1989)}]{berezhkovskii1989rate}%
  \BibitemOpen
  \bibfield  {author} {\bibinfo {author} {\bibfnamefont {A.}~\bibnamefont
  {Berezhkovskii}}, \bibinfo {author} {\bibfnamefont {L.}~\bibnamefont
  {Berezhkovskii}}, \ and\ \bibinfo {author} {\bibfnamefont {V.~Y.}\
  \bibnamefont {Zitzerman}},\ }\href@noop {} {\bibfield  {journal} {\bibinfo
  {journal} {Chemical Physics}\ }\textbf {\bibinfo {volume} {130}},\ \bibinfo
  {pages} {55} (\bibinfo {year} {1989})}\BibitemShut {NoStop}%
\bibitem [{\citenamefont {Copeland}(2016)}]{copeland2015drug}%
  \BibitemOpen
  \bibfield  {author} {\bibinfo {author} {\bibfnamefont {R.~A.}\ \bibnamefont
  {Copeland}},\ }\href@noop {} {\bibfield  {journal} {\bibinfo  {journal} {Nat.
  Rev. Drug. Discov.}\ }\textbf {\bibinfo {volume} {15}},\ \bibinfo {pages}
  {87} (\bibinfo {year} {2016})}\BibitemShut {NoStop}%
\bibitem [{\citenamefont {Tiwary}\ \emph
  {et~al.}(2015{\natexlab{a}})\citenamefont {Tiwary}, \citenamefont
  {Limongelli}, \citenamefont {Salvalaglio},\ and\ \citenamefont
  {Parrinello}}]{trypsin}%
  \BibitemOpen
  \bibfield  {author} {\bibinfo {author} {\bibfnamefont {P.}~\bibnamefont
  {Tiwary}}, \bibinfo {author} {\bibfnamefont {V.}~\bibnamefont {Limongelli}},
  \bibinfo {author} {\bibfnamefont {M.}~\bibnamefont {Salvalaglio}}, \ and\
  \bibinfo {author} {\bibfnamefont {M.}~\bibnamefont {Parrinello}},\
  }\href@noop {} {\bibfield  {journal} {\bibinfo  {journal} {Proc. Natl. Acad.
  Sci.}\ }\textbf {\bibinfo {volume} {112}},\ \bibinfo {pages} {E386} (\bibinfo
  {year} {2015}{\natexlab{a}})}\BibitemShut {NoStop}%
\bibitem [{\citenamefont {Casasnovas}\ \emph {et~al.}(2017)\citenamefont
  {Casasnovas}, \citenamefont {Limongelli}, \citenamefont {Tiwary},
  \citenamefont {Carloni},\ and\ \citenamefont {Parrinello}}]{p38}%
  \BibitemOpen
  \bibfield  {author} {\bibinfo {author} {\bibfnamefont {R.}~\bibnamefont
  {Casasnovas}}, \bibinfo {author} {\bibfnamefont {V.}~\bibnamefont
  {Limongelli}}, \bibinfo {author} {\bibfnamefont {P.}~\bibnamefont {Tiwary}},
  \bibinfo {author} {\bibfnamefont {P.}~\bibnamefont {Carloni}}, \ and\
  \bibinfo {author} {\bibfnamefont {M.}~\bibnamefont {Parrinello}},\ }\href
  {\doibase 10.1021/jacs.6b12950} {\bibfield  {journal} {\bibinfo  {journal}
  {Journal of the American Chemical Society}\ }\textbf {\bibinfo {volume}
  {139}},\ \bibinfo {pages} {4780} (\bibinfo {year} {2017})}\BibitemShut
  {NoStop}%
\bibitem [{\citenamefont {Tiwary}\ and\ \citenamefont
  {Berne}(2016{\natexlab{a}})}]{sgoop_fullerene}%
  \BibitemOpen
  \bibfield  {author} {\bibinfo {author} {\bibfnamefont {P.}~\bibnamefont
  {Tiwary}}\ and\ \bibinfo {author} {\bibfnamefont {B.~J.}\ \bibnamefont
  {Berne}},\ }\href {\doibase http://dx.doi.org/10.1063/1.4959969} {\bibfield
  {journal} {\bibinfo  {journal} {J. Chem. Phys.}\ }\textbf {\bibinfo {volume}
  {145}},\ \bibinfo {eid} {054113} (\bibinfo {year}
  {2016}{\natexlab{a}})}\BibitemShut {NoStop}%
\bibitem [{\citenamefont {Agmon}\ and\ \citenamefont
  {Rabinovich}(1992)}]{agmon1992diffusive}%
  \BibitemOpen
  \bibfield  {author} {\bibinfo {author} {\bibfnamefont {N.}~\bibnamefont
  {Agmon}}\ and\ \bibinfo {author} {\bibfnamefont {S.}~\bibnamefont
  {Rabinovich}},\ }\href@noop {} {\bibfield  {journal} {\bibinfo  {journal}
  {The Journal of chemical physics}\ }\textbf {\bibinfo {volume} {97}},\
  \bibinfo {pages} {7270} (\bibinfo {year} {1992})}\BibitemShut {NoStop}%
\bibitem [{\citenamefont {Steinbach}\ \emph {et~al.}(1991)\citenamefont
  {Steinbach}, \citenamefont {Ansari}, \citenamefont {Berendzen}, \citenamefont
  {Braunstein}, \citenamefont {Chu}, \citenamefont {Cowen}, \citenamefont
  {Ehrenstein}, \citenamefont {Frauenfelder},\ and\ \citenamefont
  {Johnson}}]{steinbach1991ligand}%
  \BibitemOpen
  \bibfield  {author} {\bibinfo {author} {\bibfnamefont {P.~J.}\ \bibnamefont
  {Steinbach}}, \bibinfo {author} {\bibfnamefont {A.}~\bibnamefont {Ansari}},
  \bibinfo {author} {\bibfnamefont {J.}~\bibnamefont {Berendzen}}, \bibinfo
  {author} {\bibfnamefont {D.}~\bibnamefont {Braunstein}}, \bibinfo {author}
  {\bibfnamefont {K.}~\bibnamefont {Chu}}, \bibinfo {author} {\bibfnamefont
  {B.~R.}\ \bibnamefont {Cowen}}, \bibinfo {author} {\bibfnamefont
  {D.}~\bibnamefont {Ehrenstein}}, \bibinfo {author} {\bibfnamefont
  {H.}~\bibnamefont {Frauenfelder}}, \ and\ \bibinfo {author} {\bibfnamefont
  {J.~B.}\ \bibnamefont {Johnson}},\ }\href@noop {} {\bibfield  {journal}
  {\bibinfo  {journal} {Biochemistry}\ }\textbf {\bibinfo {volume} {30}},\
  \bibinfo {pages} {3988} (\bibinfo {year} {1991})}\BibitemShut {NoStop}%
\bibitem [{\citenamefont {Tiwary}\ and\ \citenamefont
  {Berne}(2016{\natexlab{b}})}]{sgoop}%
  \BibitemOpen
  \bibfield  {author} {\bibinfo {author} {\bibfnamefont {P.}~\bibnamefont
  {Tiwary}}\ and\ \bibinfo {author} {\bibfnamefont {B.~J.}\ \bibnamefont
  {Berne}},\ }\href {\doibase 10.1073/pnas.1600917113} {\bibfield  {journal}
  {\bibinfo  {journal} {Proc. Natl. Acad. Sci.}\ }\textbf {\bibinfo {volume}
  {113}},\ \bibinfo {pages} {2839} (\bibinfo {year}
  {2016}{\natexlab{b}})}\BibitemShut {NoStop}%
\bibitem [{\citenamefont {Tiwary}\ \emph
  {et~al.}(2015{\natexlab{b}})\citenamefont {Tiwary}, \citenamefont {Mondal},
  \citenamefont {Morrone},\ and\ \citenamefont {Berne}}]{fullerene}%
  \BibitemOpen
  \bibfield  {author} {\bibinfo {author} {\bibfnamefont {P.}~\bibnamefont
  {Tiwary}}, \bibinfo {author} {\bibfnamefont {J.}~\bibnamefont {Mondal}},
  \bibinfo {author} {\bibfnamefont {J.~A.}\ \bibnamefont {Morrone}}, \ and\
  \bibinfo {author} {\bibfnamefont {B.~J.}\ \bibnamefont {Berne}},\ }\href@noop
  {} {\bibfield  {journal} {\bibinfo  {journal} {Proc. Natl. Acad. Sci.}\
  }\textbf {\bibinfo {volume} {112}},\ \bibinfo {pages} {12015} (\bibinfo
  {year} {2015}{\natexlab{b}})}\BibitemShut {NoStop}%
\bibitem [{\citenamefont {Press\'e}\ \emph {et~al.}(2013)\citenamefont
  {Press\'e}, \citenamefont {Ghosh}, \citenamefont {Lee},\ and\ \citenamefont
  {Dill}}]{caliber1}%
  \BibitemOpen
  \bibfield  {author} {\bibinfo {author} {\bibfnamefont {S.}~\bibnamefont
  {Press\'e}}, \bibinfo {author} {\bibfnamefont {K.}~\bibnamefont {Ghosh}},
  \bibinfo {author} {\bibfnamefont {J.}~\bibnamefont {Lee}}, \ and\ \bibinfo
  {author} {\bibfnamefont {K.~A.}\ \bibnamefont {Dill}},\ }\href {\doibase
  10.1103/RevModPhys.85.1115} {\bibfield  {journal} {\bibinfo  {journal} {Rev.
  Mod. Phys.}\ }\textbf {\bibinfo {volume} {85}},\ \bibinfo {pages} {1115}
  (\bibinfo {year} {2013})}\BibitemShut {NoStop}%
\bibitem [{\citenamefont {Dixit}\ \emph {et~al.}(2015)\citenamefont {Dixit},
  \citenamefont {Jain}, \citenamefont {Stock},\ and\ \citenamefont
  {Dill}}]{dixit2015inferring}%
  \BibitemOpen
  \bibfield  {author} {\bibinfo {author} {\bibfnamefont {P.~D.}\ \bibnamefont
  {Dixit}}, \bibinfo {author} {\bibfnamefont {A.}~\bibnamefont {Jain}},
  \bibinfo {author} {\bibfnamefont {G.}~\bibnamefont {Stock}}, \ and\ \bibinfo
  {author} {\bibfnamefont {K.~A.}\ \bibnamefont {Dill}},\ }\href@noop {}
  {\bibfield  {journal} {\bibinfo  {journal} {J. Chem. Theor. Comp.}\ }\textbf
  {\bibinfo {volume} {11}},\ \bibinfo {pages} {5464} (\bibinfo {year}
  {2015})}\BibitemShut {NoStop}%
\bibitem [{\citenamefont {Rohrdanz}\ \emph {et~al.}(2011)\citenamefont
  {Rohrdanz}, \citenamefont {Zheng}, \citenamefont {Maggioni},\ and\
  \citenamefont {Clementi}}]{diffusionmap}%
  \BibitemOpen
  \bibfield  {author} {\bibinfo {author} {\bibfnamefont {M.~A.}\ \bibnamefont
  {Rohrdanz}}, \bibinfo {author} {\bibfnamefont {W.}~\bibnamefont {Zheng}},
  \bibinfo {author} {\bibfnamefont {M.}~\bibnamefont {Maggioni}}, \ and\
  \bibinfo {author} {\bibfnamefont {C.}~\bibnamefont {Clementi}},\ }\href@noop
  {} {\bibfield  {journal} {\bibinfo  {journal} {J. Chem. Phys.}\ }\textbf
  {\bibinfo {volume} {134}},\ \bibinfo {pages} {124116} (\bibinfo {year}
  {2011})}\BibitemShut {NoStop}%
\bibitem [{\citenamefont {Coifman}\ \emph
  {et~al.}(2008{\natexlab{a}})\citenamefont {Coifman}, \citenamefont
  {Kevrekidis}, \citenamefont {Lafon}, \citenamefont {Maggioni},\ and\
  \citenamefont {Nadler}}]{coifman}%
  \BibitemOpen
  \bibfield  {author} {\bibinfo {author} {\bibfnamefont {R.~R.}\ \bibnamefont
  {Coifman}}, \bibinfo {author} {\bibfnamefont {I.~G.}\ \bibnamefont
  {Kevrekidis}}, \bibinfo {author} {\bibfnamefont {S.}~\bibnamefont {Lafon}},
  \bibinfo {author} {\bibfnamefont {M.}~\bibnamefont {Maggioni}}, \ and\
  \bibinfo {author} {\bibfnamefont {B.}~\bibnamefont {Nadler}},\ }\href@noop {}
  {\bibfield  {journal} {\bibinfo  {journal} {Mult. Mod. Sim.}\ }\textbf
  {\bibinfo {volume} {7}},\ \bibinfo {pages} {842} (\bibinfo {year}
  {2008}{\natexlab{a}})}\BibitemShut {NoStop}%
\bibitem [{\citenamefont {Coifman}\ \emph
  {et~al.}(2008{\natexlab{b}})\citenamefont {Coifman}, \citenamefont
  {Kevrekidis}, \citenamefont {Lafon}, \citenamefont {Maggioni},\ and\
  \citenamefont {Nadler}}]{coifman2008diffusion}%
  \BibitemOpen
  \bibfield  {author} {\bibinfo {author} {\bibfnamefont {R.~R.}\ \bibnamefont
  {Coifman}}, \bibinfo {author} {\bibfnamefont {I.~G.}\ \bibnamefont
  {Kevrekidis}}, \bibinfo {author} {\bibfnamefont {S.}~\bibnamefont {Lafon}},
  \bibinfo {author} {\bibfnamefont {M.}~\bibnamefont {Maggioni}}, \ and\
  \bibinfo {author} {\bibfnamefont {B.}~\bibnamefont {Nadler}},\ }\href@noop {}
  {\bibfield  {journal} {\bibinfo  {journal} {Mult. Mod. Sim.}\ }\textbf
  {\bibinfo {volume} {7}},\ \bibinfo {pages} {842} (\bibinfo {year}
  {2008}{\natexlab{b}})}\BibitemShut {NoStop}%
\bibitem [{\citenamefont {Bicout}\ and\ \citenamefont
  {Szabo}(1998)}]{szabo_bicout}%
  \BibitemOpen
  \bibfield  {author} {\bibinfo {author} {\bibfnamefont {D.}~\bibnamefont
  {Bicout}}\ and\ \bibinfo {author} {\bibfnamefont {A.}~\bibnamefont {Szabo}},\
  }\href@noop {} {\bibfield  {journal} {\bibinfo  {journal} {J. Chem. Phys.}\
  }\textbf {\bibinfo {volume} {109}},\ \bibinfo {pages} {2325} (\bibinfo {year}
  {1998})}\BibitemShut {NoStop}%
\bibitem [{\citenamefont {Straub}, \citenamefont {Berne},\ and\ \citenamefont
  {Roux}(1990)}]{straub1990spatial}%
  \BibitemOpen
  \bibfield  {author} {\bibinfo {author} {\bibfnamefont {J.~E.}\ \bibnamefont
  {Straub}}, \bibinfo {author} {\bibfnamefont {B.~J.}\ \bibnamefont {Berne}}, \
  and\ \bibinfo {author} {\bibfnamefont {B.}~\bibnamefont {Roux}},\ }\href@noop
  {} {\bibfield  {journal} {\bibinfo  {journal} {The Journal of chemical
  physics}\ }\textbf {\bibinfo {volume} {93}},\ \bibinfo {pages} {6804}
  (\bibinfo {year} {1990})}\BibitemShut {NoStop}%
\bibitem [{\citenamefont {Hummer}(2005)}]{hummer2005position}%
  \BibitemOpen
  \bibfield  {author} {\bibinfo {author} {\bibfnamefont {G.}~\bibnamefont
  {Hummer}},\ }\href@noop {} {\bibfield  {journal} {\bibinfo  {journal} {N.
  Jour. Phys.}\ }\textbf {\bibinfo {volume} {7}},\ \bibinfo {pages} {34}
  (\bibinfo {year} {2005})}\BibitemShut {NoStop}%
\bibitem [{\citenamefont {Sohl-Dickstein}, \citenamefont {Battaglino},\ and\
  \citenamefont {DeWeese}(2011)}]{jascha_prl}%
  \BibitemOpen
  \bibfield  {author} {\bibinfo {author} {\bibfnamefont {J.}~\bibnamefont
  {Sohl-Dickstein}}, \bibinfo {author} {\bibfnamefont {P.~B.}\ \bibnamefont
  {Battaglino}}, \ and\ \bibinfo {author} {\bibfnamefont {M.~R.}\ \bibnamefont
  {DeWeese}},\ }\href@noop {} {\bibfield  {journal} {\bibinfo  {journal}
  {Physical review letters}\ }\textbf {\bibinfo {volume} {107}},\ \bibinfo
  {pages} {220601} (\bibinfo {year} {2011})}\BibitemShut {NoStop}%
\bibitem [{\citenamefont {Sohl-Dickstein}, \citenamefont {Battaglino},\ and\
  \citenamefont {DeWeese}(2009)}]{jascha_picml}%
  \BibitemOpen
  \bibfield  {author} {\bibinfo {author} {\bibfnamefont {J.}~\bibnamefont
  {Sohl-Dickstein}}, \bibinfo {author} {\bibfnamefont {P.}~\bibnamefont
  {Battaglino}}, \ and\ \bibinfo {author} {\bibfnamefont {M.~R.}\ \bibnamefont
  {DeWeese}},\ }\href@noop {} {\bibfield  {journal} {\bibinfo  {journal}
  {Proceedings of thw 28th International Conference on Machine
  Learning;arXiv:0906.4779}\ } (\bibinfo {year} {2009})}\BibitemShut {NoStop}%
\bibitem [{\citenamefont {Tiwary}\ and\ \citenamefont
  {Berne}(2016{\natexlab{c}})}]{kramers}%
  \BibitemOpen
  \bibfield  {author} {\bibinfo {author} {\bibfnamefont {P.}~\bibnamefont
  {Tiwary}}\ and\ \bibinfo {author} {\bibfnamefont {B.~J.}\ \bibnamefont
  {Berne}},\ }\href@noop {} {\bibfield  {journal} {\bibinfo  {journal} {J.
  Chem. Phys.}\ }\textbf {\bibinfo {volume} {144}},\ \bibinfo {pages} {134103}
  (\bibinfo {year} {2016}{\natexlab{c}})}\BibitemShut {NoStop}%
\bibitem [{\citenamefont {Tiwary}\ and\ \citenamefont
  {Parrinello}(2014)}]{tiwary_rewt}%
  \BibitemOpen
  \bibfield  {author} {\bibinfo {author} {\bibfnamefont {P.}~\bibnamefont
  {Tiwary}}\ and\ \bibinfo {author} {\bibfnamefont {M.}~\bibnamefont
  {Parrinello}},\ }\href@noop {} {\bibfield  {journal} {\bibinfo  {journal} {J.
  Phys. Chem. B}\ }\textbf {\bibinfo {volume} {119}},\ \bibinfo {pages} {736}
  (\bibinfo {year} {2014})}\BibitemShut {NoStop}%
\bibitem [{\citenamefont {Berezhkovskii}\ and\ \citenamefont
  {Szabo}(2013)}]{berezhkovskii2013diffusion}%
  \BibitemOpen
  \bibfield  {author} {\bibinfo {author} {\bibfnamefont {A.~M.}\ \bibnamefont
  {Berezhkovskii}}\ and\ \bibinfo {author} {\bibfnamefont {A.}~\bibnamefont
  {Szabo}},\ }\href@noop {} {\bibfield  {journal} {\bibinfo  {journal} {The
  journal of physical chemistry. B}\ }\textbf {\bibinfo {volume} {117}},\
  \bibinfo {pages} {13115} (\bibinfo {year} {2013})}\BibitemShut {NoStop}%
\bibitem [{\citenamefont {De~Leon}\ and\ \citenamefont
  {Berne}(1981)}]{deleonberne}%
  \BibitemOpen
  \bibfield  {author} {\bibinfo {author} {\bibfnamefont {N.}~\bibnamefont
  {De~Leon}}\ and\ \bibinfo {author} {\bibfnamefont {B.}~\bibnamefont
  {Berne}},\ }\href@noop {} {\bibfield  {journal} {\bibinfo  {journal} {J.
  Chem. Phys.}\ }\textbf {\bibinfo {volume} {75}},\ \bibinfo {pages} {3495}
  (\bibinfo {year} {1981})}\BibitemShut {NoStop}%
\bibitem [{\citenamefont {Bussi}\ and\ \citenamefont
  {Parrinello}(2007)}]{bussi_langevin}%
  \BibitemOpen
  \bibfield  {author} {\bibinfo {author} {\bibfnamefont {G.}~\bibnamefont
  {Bussi}}\ and\ \bibinfo {author} {\bibfnamefont {M.}~\bibnamefont
  {Parrinello}},\ }\href@noop {} {\bibfield  {journal} {\bibinfo  {journal}
  {Phys. Rev. E}\ }\textbf {\bibinfo {volume} {75}},\ \bibinfo {pages} {056707}
  (\bibinfo {year} {2007})}\BibitemShut {NoStop}%
\bibitem [{\citenamefont {Poon}\ and\ \citenamefont
  {Peters}(2015)}]{poon2015accelerated}%
  \BibitemOpen
  \bibfield  {author} {\bibinfo {author} {\bibfnamefont {G.~G.}\ \bibnamefont
  {Poon}}\ and\ \bibinfo {author} {\bibfnamefont {B.}~\bibnamefont {Peters}},\
  }\href@noop {} {\bibfield  {journal} {\bibinfo  {journal} {The Journal of
  Physical Chemistry B}\ }\textbf {\bibinfo {volume} {120}},\ \bibinfo {pages}
  {1679} (\bibinfo {year} {2015})}\BibitemShut {NoStop}%
\bibitem [{\citenamefont {Valsson}, \citenamefont {Tiwary},\ and\ \citenamefont
  {Parrinello}(2016)}]{arpc_meta}%
  \BibitemOpen
  \bibfield  {author} {\bibinfo {author} {\bibfnamefont {O.}~\bibnamefont
  {Valsson}}, \bibinfo {author} {\bibfnamefont {P.}~\bibnamefont {Tiwary}}, \
  and\ \bibinfo {author} {\bibfnamefont {M.}~\bibnamefont {Parrinello}},\
  }\href@noop {} {\bibfield  {journal} {\bibinfo  {journal} {Ann. Rev. Phys.
  Chem.}\ }\textbf {\bibinfo {volume} {67}},\ \bibinfo {pages} {159} (\bibinfo
  {year} {2016})}\BibitemShut {NoStop}%
\bibitem [{\citenamefont {Tiwary}\ and\ \citenamefont {van~de
  Walle}(2016)}]{tiwary2016review}%
  \BibitemOpen
  \bibfield  {author} {\bibinfo {author} {\bibfnamefont {P.}~\bibnamefont
  {Tiwary}}\ and\ \bibinfo {author} {\bibfnamefont {A.}~\bibnamefont {van~de
  Walle}},\ }in\ \href@noop {} {\emph {\bibinfo {booktitle} {Multiscale
  Materials Modeling for Nanomechanics}}}\ (\bibinfo  {publisher} {Springer},\
  \bibinfo {year} {2016})\ pp.\ \bibinfo {pages} {195--221}\BibitemShut
  {NoStop}%
\bibitem [{\citenamefont {McCarty}\ and\ \citenamefont
  {Parrinello}(2017)}]{mccarty2017variational}%
  \BibitemOpen
  \bibfield  {author} {\bibinfo {author} {\bibfnamefont {J.}~\bibnamefont
  {McCarty}}\ and\ \bibinfo {author} {\bibfnamefont {M.}~\bibnamefont
  {Parrinello}},\ }\href@noop {} {\bibfield  {journal} {\bibinfo  {journal}
  {arXiv preprint arXiv:1703.08777}\ } (\bibinfo {year} {2017})}\BibitemShut
  {NoStop}%
\end{thebibliography}
	\end{document}